\documentclass[sigconf, nonacm]{acmart}
\setcopyright{none}

\newcommand{\artifacturl}{https://github.com/wrp-wrp/WaRP}

\AtBeginDocument{%
  }

\usepackage{algorithm}
\usepackage[noend]{algpseudocode}
\usepackage{dblfloatfix}
\usepackage{placeins}
\usepackage{multirow}
\usepackage{siunitx}
\usepackage[noabbrev,capitalize,nameinlink]{cleveref}
\usepackage{subcaption}

\graphicspath{{figures/}}

\newcommand{\secref}[1]{~\cref{#1}}
\newcommand{\figref}[1]{~\cref{#1}}
\newcommand{\tabref}[1]{~\cref{#1}}

\newcommand{\system}{WaRP}
\usepackage{layout-preservation}
\DeclareMathOperator*{\argmin}{arg\,min}
\algrenewcommand\algorithmicrequire{\textbf{Input:}}
\algrenewcommand\algorithmicensure{\textbf{Output:}}

\begin{document}
\flushbottom
\setlength{\emergencystretch}{3em}

\title[\system{}]{\system{}: Workload-Aware Request Planning for Vector Retrieval on Object Storage}

\settopmatter{authorsperrow=2,printfolios=true}
\author{Ruipeng Wan}
\affiliation[obeypunctuation=true]{%
  \institution{Huazhong University of Science and Technology}\country{}}
\email{rprp@hust.edu.cn}

\author{Xinyu Zeng}
\affiliation[obeypunctuation=true]{%
  \institution{Modalink}\country{}}
\email{xinyu@modalink.ai}

\author{Yiming Qiao}
\affiliation[obeypunctuation=true]{%
  \institution{Tsinghua University}\country{}}
\email{qiaoym21@mails.tsinghua.edu.cn}

\author{Huanchen Zhang$^{\ast}$}
\affiliation[obeypunctuation=true]{%
  \institution{Tsinghua University}\country{}}
\email{huanchen@tsinghua.edu.cn}

\begin{abstract}
Modern vector search systems keep payload vectors in cloud object storage and
only a compact routing index on fast local storage, typically using
cluster-based ANN indexes. This design inherits a problem existing systems
do not address. On object storage, small Range-GETs have latency
dominated by fixed per-request overhead and are billed per request.
Existing readers either issue one request per probed cluster, or
coalesce adjacent ranges with hand-tuned gap or span thresholds that do
not transfer across datasets or query workloads.\looseness=-1

We formulate the grouping of probed-cluster byte intervals into object-store
requests as a global per-query optimization problem and present \system{}, a
workload-aware request planner that solves it efficiently. Given the sorted
byte intervals, a limit on the number of requests per query, and an estimate
of how many object-store requests from one query can proceed concurrently
under the target serving configuration, \system{} searches for a merge plan
that minimizes a concurrency-aware latency surrogate while satisfying the
request limit. Using an optimized dynamic program and adaptive frontier search,
\system{} returns a plan within $(1{+}\epsilon)$ of the optimum, with per-query
planning overhead under $10$\,ms even in the most demanding tested settings. A
lightweight offline reordering pass further improves request mergeability
without changing the semantics of ANN search.

We implement \system{} inside Lance's I/O scheduler and evaluate it on
AWS S3. We build IVFPQ indexes on seven datasets, including SIFT100M
and DEEP100M at the 100M-vector scale. \system{} + Reorder improves QPS
by $1.09\text{--}3.48\times$ and reduces GETs by
$2.12\text{--}5.19\times$ relative to Lance, while preserving Lance's
recall.

\end{abstract}

\settopmatter{printccs=false}

\maketitle

\pagestyle{warppreprint}
\thispagestyle{warppreprint}
\newcommand{\publicationreference}{%
\begingroup\small\noindent\raggedright\textbf{PVLDB Reference Format:}\\
Ruipeng Wan, Xinyu Zeng, Yiming Qiao, Huanchen Zhang. \shorttitle.\\
\layouttextline{5.644pt}{1.616pt}{PVLDB, Vol.~20. Accept with Availability Check.}\\
\layouttextline{5.496pt}{0.472pt}{Publication details to follow.}
\endgroup}
\begingroup\small\noindent\raggedright
\phantom{\textbf{PVLDB Reference Format:}}\\
\phantom{Ruipeng Wan$^{\dagger}$, Xinyu Zeng$^{\ddagger}$, Yiming Qiao, Huanchen Zhang$^{\ast}$. \shorttitle.}\\
\layoutreserve{0pt}{5.644pt}{1.616pt}\\
\layoutreserve{0pt}{5.496pt}{0.472pt}
\endgroup
\begingroup
\renewcommand\thefootnote{}\footnote{\noindent
\phantom{$^{\ast}$Corresponding author.}\\
\layoutreserve{0pt}{4.80899pt}{0.08049pt}\\
\layoutreserve{0pt}{4.80899pt}{1.652pt}\\
\layoutreserve{0pt}{4.80899pt}{1.652pt}\\
\layoutreserve{0pt}{4.93849pt}{1.652pt}\\
\layoutreserve{0pt}{4.80899pt}{0.08049pt}\\
\raggedright\layoutreserve{0pt}{4.80899pt}{1.652pt}\\
\layoutreserve{0pt}{4.80899pt}{0.413pt}\\
}\addtocounter{footnote}{-1}\endgroup

\newcommand{\publicationfooter}{%
$^{\ast}$Corresponding author.\\
This work is licensed under the Creative Commons BY-NC-ND 4.0 International License. Visit \url{https://creativecommons.org/licenses/by-nc-nd/4.0/} to view a copy of this license. For any use beyond those covered by this license, obtain permission by emailing \href{mailto:info@vldb.org}{info@vldb.org}. Copyright is held by the owner/author(s). Publication rights licensed to the VLDB Endowment. \\
\raggedright Proceedings of the VLDB Endowment, Vol.~20.\\
Final version details to follow.\\
}
\newcommand{\publicationartifact}{%
\begingroup\small\noindent\raggedright\textbf{PVLDB Artifact Availability:}\\
The source code, data, and/or other artifacts have been made available at \url{\artifacturl}.
\endgroup}
\AddToShipoutPictureFG*{\AtPageUpperLeft{%
  \put(53.798bp,-582bp){\parbox[t]{\columnwidth}{\normalfont\publicationreference}}%
  \put(317.948bp,-379bp){\parbox[t]{\columnwidth}{\normalfont\publicationartifact}}%
  \put(53.798bp,-640bp){\rule{4pc}{0.4pt}}%
  \put(53.798bp,-650.19bp){\parbox[t]{\columnwidth}{\normalfont\footnotesize\noindent\publicationfooter}}%
}}

\ifdefempty{\artifacturl}{}{
\vspace{.3cm}
\begingroup\small\noindent\raggedright
\phantom{\textbf{PVLDB Artifact Availability:}}\\[-31.37521pt]
\phantom{The source code, data, and/or other artifacts have been made available at}\\
\phantom{\url{\artifacturl}.}
\endgroup
}

\section{Introduction}
\label{sec:intro}

Large-scale vector search is increasingly deployed over disaggregated storage,
with payload vectors placed in cloud object stores to reduce storage cost and
decouple storage from compute~\cite{verbitski2017amazon_aurora,dageville2016snowflake,pinecone2024slab,databricks2025decoupled,guo2022manucloudnativevector,turbopuffer2026architecture}.
In these systems, a compact routing structure remains on fast local storage,
while query execution first selects a probe set and then fetches the
corresponding payloads remotely. This separation is well suited to
cluster-based index families such as IVF~\cite{faissivf} and
IVFPQ~\cite{ivfPQsift}, and to systems built on them such as
SPFresh~\cite{xu2023spfresh}. Once probing finishes, the system knows
which cluster payloads are required, but it must still decide how to translate
their byte intervals into remote requests.

\begin{figure}[!tbp]
  \setlength{\abovecaptionskip}{2pt}
  \setlength{\belowcaptionskip}{0pt}
  \centering
  \includegraphics[width=\linewidth]{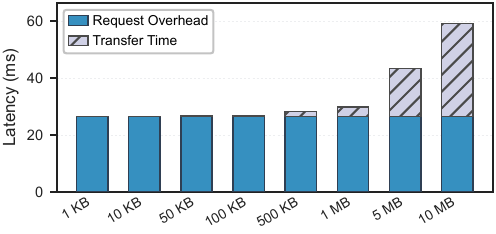}
  \caption{\textbf{S3 Range-GET latency.} Per-request latency decomposed into time-to-first-byte and transfer time across payload sizes.}
  \Description{A latency decomposition plot for S3 range requests across payload sizes, showing that small requests are dominated by fixed overhead rather than transfer time.}
  \label{fig:intro-latency-composition}
\end{figure}

\begin{figure}[!tbp]
  \setlength{\abovecaptionskip}{2pt}
  \setlength{\belowcaptionskip}{9pt}
  \centering
  \layoutmarginfloat{27.72894pt}{5.98949pt}{2.64598pt}%
  \includegraphics[width=\columnwidth]{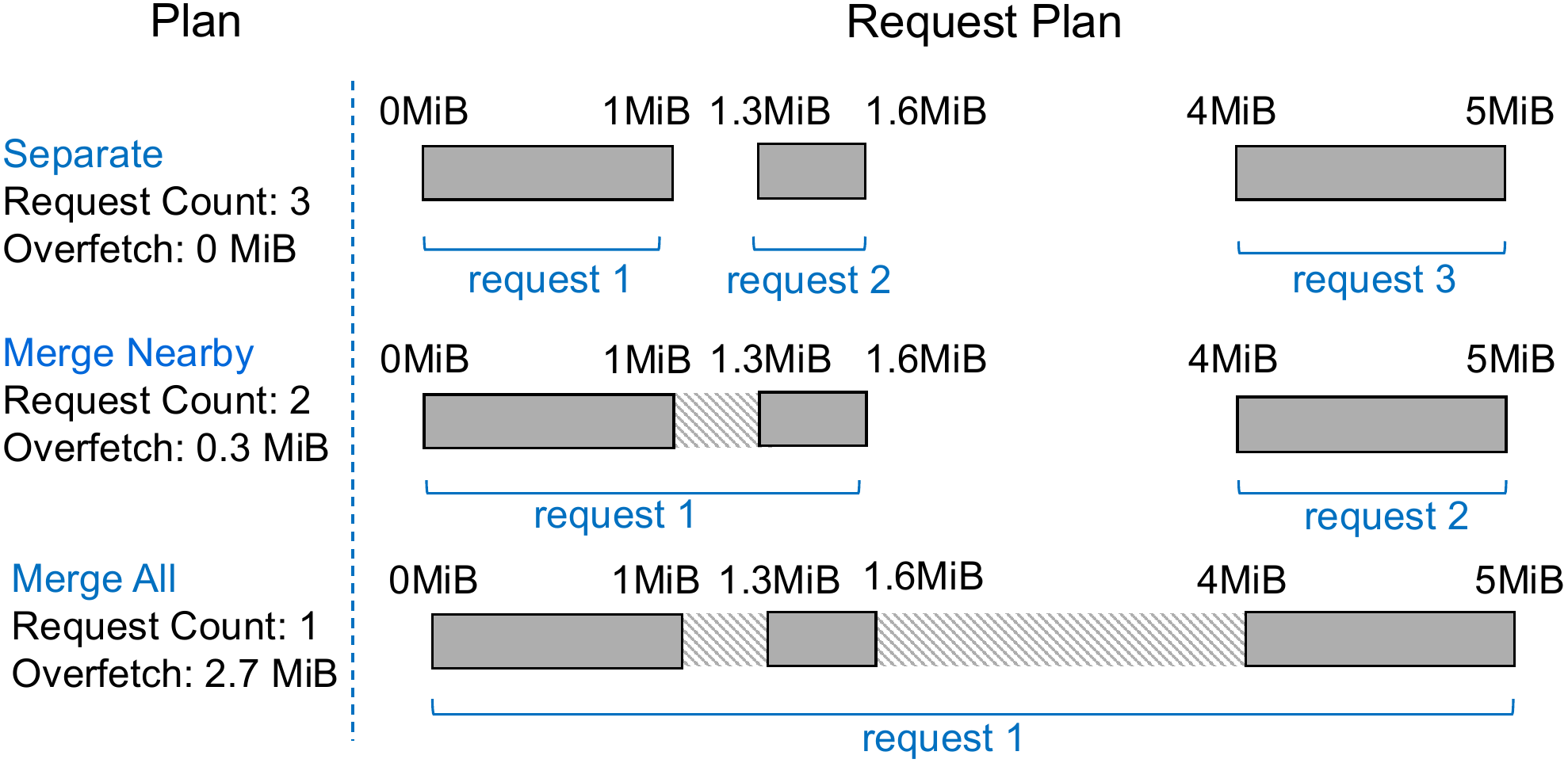}%
  \caption{\textbf{Request plans for three probed cluster payloads.}}
  \Description{Three request plans for the same three cluster payloads, showing that merging reduces the number of Range-GETs while increasing the number of overfetched bytes.}
  \label{fig:intro-plan-example}
\end{figure}

\layoutmargin{27.72894pt}{5.98949pt}{2.64598pt}
\papertext{Object-store Range-GETs have much higher fixed overhead than local
reads and exhibit high tail latency. As \figref{fig:intro-latency-composition}
shows, a 1~KB request takes roughly 25~ms, while a 1~MB request takes only
about 30~ms. Small cluster reads are therefore dominated by per-request
overhead. Object stores also charge per API request, so fragmented retrieval
increases both latency and GET charges.
\figref{fig:intro-plan-example} makes the resulting choice concrete. Probe
selection has picked three cluster payloads in the same object. Reading them
separately takes three Range-GETs. Merging the first two saves one request and
overfetches the 0.3~MiB gap between them. Merging all three leaves a single
request but overfetches 2.7~MiB, because the last payload sits 2.4~MiB away.
Probe selection determines the required cluster payloads. Request planning
groups their byte intervals to balance per-request overhead against overfetch.}

Existing systems typically handle this stage with local coalescing heuristics,
such as greedy merging with fixed gap
thresholds~\cite{lance,databricks2025decoupled,ZengHSPMZ23}. Such hand-tuned rules are hard
to rely on because the best thresholds vary across datasets and query
workloads. We confirm this empirically
in\secref{sec:eval-objective}. Careful tuning works well on one dataset, but
performance drops sharply when the same parameters are applied to another.
Request planning is therefore better treated as a global optimization problem
over the ordered payload intervals exposed after probe selection.

Another way to reduce per-request overhead is to cluster more
coarsely when building the ANN index. Each remote read then
covers more payload and amortizes the fixed overhead over a
larger transfer. But this only shifts cost from I/O to
compute and index maintenance without changing how requests are
formed. In IVF terms, the index has fewer, larger clusters and the
query probes fewer of them. As shown in~\tabref{tab:intro-coarse-cluster},
at matched recall the coarser setting cuts request count by 58\%
but enlarges the candidate set by nearly 50\%, raising per-query
compute. Larger clusters also weaken pruning, make updates and
rewrites costlier, and limit flexibility under filters or mixed
workloads. Tuning probe granularity is therefore not a substitute
for solving request planning directly.

We propose \system{}, a \textbf{w}orkload-\textbf{a}ware online
\textbf{r}equest \textbf{p}lanner that addresses this trade-off, and pair it
with an offline method that reorders IVF cluster payloads specifically for
object-store request planning.
Given a request budget, an estimate of how many object-store requests from one
query can proceed concurrently under the target serving configuration, and a
request-time surrogate calibrated offline from object-store measurements,
\system{} returns a merge plan provably within
$(1{+}\epsilon)$ of the optimum, with per-query overhead under $10$\,ms
even in the most demanding tested settings. \layoutmargin{27.72894pt}{5.98949pt}{2.64598pt}
\papertext{By bounding the number of GETs per query, the request budget directly
bounds the query's GET charge and provides a single knob for balancing that
charge against latency.} The reordering method places likely co-accessed clusters nearby,
shrinking the gaps that merged reads must bridge and reducing overfetch.
Request planning and reordering preserve probe selection and recall.
For a fixed probe set, all methods retrieve the same clusters and differ only
in how the selected payloads are laid out and fetched.

We implement \system{} inside Lance's I/O scheduler~\cite{lance}
and pair it with offline reordering of IVF cluster payloads. Across seven
datasets on AWS S3~\cite{amazonS3}, \system{} + Reorder delivers
$1.09\text{--}3.48\times$ the QPS of Lance and a
$2.12\text{--}5.19\times$ GET reduction while preserving Lance's recall.

\papertext{This paper makes four contributions. First, we identify request planning
as a missing optimization layer between probe selection and payload retrieval
in object-storage-backed ANN search. We formulate it as a global per-query
optimization problem.
Second, we make this optimization practical with an optimized dynamic
program and adaptive frontier search. The resulting planner is fast enough to
run on the query critical path.
Third, we introduce offline reordering of IVF cluster payloads for
object-store request planning, creating better merge opportunities
with less overfetch while preserving probe selection and recall.
Finally, we implement \system{} in Lance and evaluate its end-to-end
performance across seven datasets on AWS S3.
}
\layoutmargin{25.52394pt}{15.10497pt}{2.64598pt}

\begin{table}[!tbp]
\caption{\textbf{Recall-Matched Comparison of Fine vs.\ Coarse Clustering on SIFT1M.} Coarser clustering reduces requests but enlarges the candidate set, raising per-query compute.}
\label{tab:intro-coarse-cluster}
\centering
\setlength{\tabcolsep}{4pt}
\renewcommand{\arraystretch}{1.15}
\resizebox{\columnwidth}{!}{%
\begin{tabular}{lccccc}
\toprule
Setting & Lists & Probes & Recall@10 & GET/q & Cand./q \\
\midrule
Fine clustering & 4096 & 192 & 0.9973 & 192 & 57.4K \\
Coarse clustering & 1024 & 80 & 0.9970 & 80 & 85.0K \\
\bottomrule
\end{tabular}%
}
\end{table}

\begin{figure*}[!tbp]
  \centering
  \includegraphics[width=0.88\textwidth]{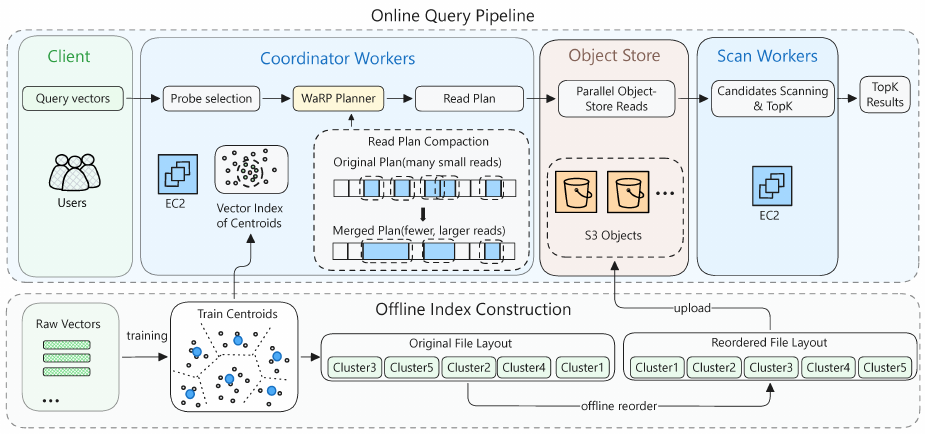}
  \caption{\textbf{\system{} in the query execution pipeline.} Online, the request planner runs after probe selection and before object-store reads. Offline, a reordering pass tightens inter-cluster gaps.}
  \Description{Pipeline diagram of \system{} with query submission, probe selection, interval generation, request planning, merged range-get execution, and reranking, plus an offline reordering pass that tightens inter-cluster gaps.}
  \label{fig:warp-pipeline}
\end{figure*}

\section{Background and Motivation}
\label{sec:background}

In this section, we discuss cluster-based ANN indexes and the I/O characteristics of cloud object storage. We also describe the execution pipeline of cloud-based ANN queries.

\subsection{Cluster-Based ANN Indexes}
\label{sec:bg-index}

Cluster-based ANN indexes, such as IVF~\cite{faissivf}, partition
vectors into clusters. To answer a query, the system first selects a
probe set by comparing the query to cluster centroids, then fetches
the payloads of those clusters for candidate evaluation.
We use cluster-based ANN indexes broadly to include plain IVF and its variants,
such as IVFPQ~\cite{ivfPQsift} and SPFresh~\cite{xu2023spfresh}.
We also include designs that build an auxiliary graph or tree over the
cluster centroids---using HNSW~\cite{HNSW}, Vamana~\cite{diskann},
NSG~\cite{NSG}, or a KD-tree~\cite{KDTree}---to accelerate probe-set
selection. These auxiliary structures change how clusters are selected,
not how their payloads are retrieved. All such methods therefore remain
in the cluster-based setting considered here.

\subsection{Cloud Object-Store I/O Characteristics}
\label{sec:bg-storage}

Cloud object stores provide low-cost, elastic capacity for ANN index payloads.
Representative services include Amazon S3, Azure Blob Storage, and Google
Cloud Storage~\cite{amazonS3,azureBlobStorage,googleCloudStorage}. Two aspects
of their I/O behavior matter here. Each Range-GET carries substantial fixed
overhead, and request latency is highly variable with a long tail.

\begin{figure}[!tbp]
  \centering
  \layoutmarginfloat{25.52394pt}{5.98949pt}{2.64598pt}%
  \begin{subfigure}[t]{0.66\linewidth}
    \centering
    \includegraphics[width=\linewidth]{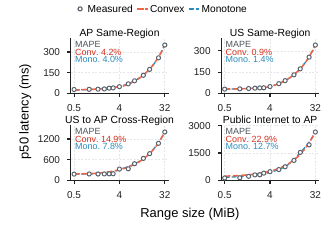}
    \caption{\papertext{Surrogate fits across four deployment paths.}}
    \Description{Four panels compare measured single-request p50 latency
    with convex two-segment and monotone piecewise-linear fits for two
    same-region paths, a cross-region path, and a public-Internet path.
    Each panel reports held-out mean absolute percentage error.}
    \label{fig:bg-calibration}
  \end{subfigure}\hfill
  \begin{subfigure}[t]{0.32\linewidth}
    \centering
    \includegraphics[width=\linewidth]{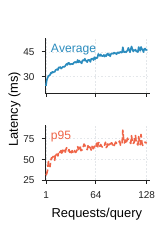}
    \caption{\papertext{High tail latency.}}
    \Description{Average and p95 completion time versus requests per query for tiny real-S3 reads.}
    \label{fig:bg-request-count-tail}
  \end{subfigure}
  \caption{\textbf{Object-store I/O on AWS S3.} }
  \Description{Two plots shown side by side. Left: measured
  single-request p50 latency with convex and monotone surrogate fits
  across four deployment paths. Right: average and p95 completion time
  versus requests per query for tiny real-S3 reads.}
  \label{fig:bg-microbenchmarks}
\end{figure}

The latency decomposition in\figref{fig:intro-latency-composition} shows
the fixed overhead of each Range-GET, while\figref{fig:bg-request-count-tail}
shows that both average and p95 query latency rise markedly as more 1~KB
requests are issued in parallel. Merging requests reduces this fan-out but
may overfetch data, creating a trade-off between latency, request count, and
transferred bytes.
\layoutmargin{27.72894pt}{15.10497pt}{2.64598pt}
\papertext{Request count also determines GET charge. In the common same-region
setting, data transfer is not billed~\cite{amazonS3Pricing}, so GET requests
are the primary monetary cost. Where transferred bytes are billed, overfetch
also incurs data-transfer charges.}

\papertext{\Cref{fig:bg-calibration} shows that measured single-request
latency across four deployment paths can be approximated by simple
piecewise-linear functions.\layoutmargin{25.52394pt}{5.98949pt}{2.64598pt} The figure compares
two such fits. The
``Convex'' fit is monotone and convex, while the more general ``Monotone''
fit only requires monotonicity. It also reports their mean absolute
percentage error (MAPE) on request sizes excluded from fitting. \system{}
uses the calibrated fit as the request-time surrogate when planning each
query.}

\subsection{Execution Pipeline and Planning Challenge}
\label{sec:bg-cloud-ann}

\layoutmargin{27.72894pt}{24.22044pt}{2.64598pt}
\papertext{As illustrated in\figref{fig:warp-pipeline}, we target deployments
in which cluster payloads remain in object storage. For each query, we use a
local routing structure to select the nearest cluster centroids, then fetch
payloads that miss the local cache from object storage. Even when an index fits
in local memory, loading the entire index from object storage before serving
queries can be slow, while a QueryNode serving multiple datasets may not have
enough local capacity to keep every index resident. The resulting pipeline
exposes a key step that has received little attention in prior work: deciding
how to turn the required payload accesses into actual object-store requests.
We call this step \emph{request planning}. A direct implementation issues one
Range-GET per cache-missing cluster, so its request count grows with cache
misses. Request planning instead decides which nearby reads to
merge into fewer requests. Optimizing this step is the focus of \system{}.}
After the planned Range-GETs complete, candidate evaluation or reranking
proceeds.

\section{Problem Formulation}
\label{sec:formulation}

\layoutmargin{27.72894pt}{5.98949pt}{2.64598pt}
\papertext{For each query, request planning groups the required payload intervals
into one merge plan while preserving every required payload. Under a
fixed probe set and evaluation procedure, the candidate set and recall
therefore remain unchanged.}

\textit{Input.}
\papertext{Given required intervals
$\mathcal{I}=\{[l_i,r_i)\}_{i=1}^{n}$ with $r_i \le l_{i+1}$ for
$1 \le i < n$, and a request budget $K_{\max}$,
request planning produces at most $K_{\max}$ planned ranges. Each becomes one payload GET, so this
budget directly bounds the query's GET charge.}

The planner scores a candidate range through a request-time surrogate
$\tau(\ell)$, which maps planned range length $\ell$ to the estimated
completion time of one Range-GET. We assume $\tau$ is nondecreasing,
as every fit in\figref{fig:bg-calibration} is.

\textit{Plan.} A feasible merge plan for $\mathcal{I}$ partitions it into
$K$ contiguous groups, with $1 \le K \le K_{\max}$. Each
group becomes one planned range $[L_u, R_u)$, with $L_u$ the left
endpoint of the first interval in the group and $R_u$ the right
endpoint of the last. The plan is the ordered sequence
$\pi = ([L_1, R_1), \dots, [L_K, R_K))$. We write $K(\pi)=K$ for the
planned range count and $\ell_u = R_u - L_u$ for the byte length of
planned range $u$, which may include gaps between required intervals assigned to
the same merge.

\textit{Objective.} The quality of a plan $\pi$ depends on two quantities:
\begin{align*}
 W(\pi) &= \sum_{u=1}^{K(\pi)} \tau(\ell_u), \\
 B(\pi) &= \max_{1 \le u \le K(\pi)} \tau(\ell_u).
\end{align*}
$W(\pi)$ is the sum of request-time estimates over all planned ranges.
It is the modeled completion time at the serial endpoint, where the
$K(\pi)$ requests are issued one at a time. $B(\pi)$ is the maximum
request-time estimate among the planned ranges. It is the modeled
completion time at the ideal-parallel endpoint, where all requests
start together and the query waits for the slowest one.

\layoutmargin{27.72894pt}{5.98949pt}{2.64598pt}
\papertext{$W(\pi)$ and $B(\pi)$ capture the serial and ideal-parallel
extremes. For systems between them, the request parallelism estimate
$P$ describes how many payload GETs from one query can progress
concurrently. Using Graham's bound~\cite{graham1969boundsdiaodu}, we
define}
\begin{equation*}
T_P(\pi)
= \frac{W(\pi)}{P} + \left(1-\frac{1}{P}\right)B(\pi).
\end{equation*}
\papertext{$T_P$ is a planning objective, not an exact latency predictor.
It reduces to $W(\pi)$ at $P=1$ and approaches $B(\pi)$ as
$P\to\infty$. At planning time, $P$ reflects the request parallelism
expected under the target serving configuration.} \\

\noindent\fbox{\parbox{\dimexpr\linewidth-2\fboxsep-2\fboxrule\relax}{%
\textbf{Request Planning Problem.}\\[3pt]
\textbf{Input:} a sorted interval set $\mathcal{I}$, a request
budget $K_{\max}$, and a request parallelism estimate $P$.\\
\textbf{Output:} a feasible plan $\pi^{*}$ minimizing
\begin{equation*}
T(\pi)=
\begin{cases}
W(\pi), & \text{serial execution},\\
B(\pi), & \text{ideal parallel execution},\\
T_P(\pi), & \text{finite-concurrency execution}.
\end{cases}
\end{equation*}
}}
\\

All three objectives depend on a plan only through
$(W(\pi), B(\pi))$, and none of them increases when either metric
decreases. The Pareto frontier in the $(B,W)$ space therefore
contains an \mbox{optimal plan.}

\section{Request Planning}
\label{sec:planning}

\begin{figure}[!tbp]
  \centering
  \includegraphics[width=\columnwidth]{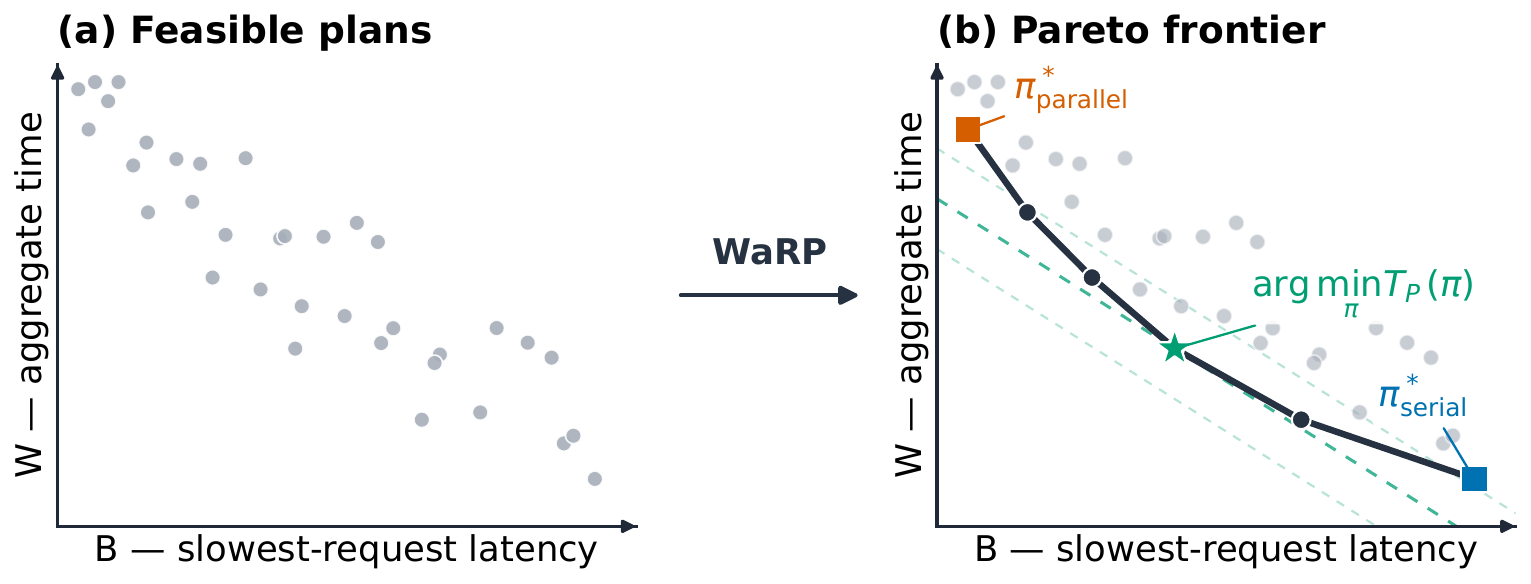}
  \caption{\textbf{From design space to Pareto frontier.}
  \emph{Left}: feasible merge plans for one query in the $(B, W)$
  plane. \emph{Right}: their lower convex hull, with corners at the
  $\min W$ and $\min B$ plans. The finite-$P$ optimum (star) is the
  lowest-$T_P$ feasible vertex.}
  \Description{Two side-by-side panels sharing axes B (maximum request-time
  estimate, x) and W (sum of request-time estimates, y). The left panel shows grey
  points. The right panel shows the same points with their lower convex hull
  drawn through them and three points highlighted: a red square at the
  upper-left for the minimum-B plan, a blue square at the lower-right for
  the minimum-W plan, and a green star in the middle for the finite-P
  optimum, with a dashed green tangent line. An arrow
  labelled WaRP connects the two panels.}
  \label{fig:planning-frontier}
\end{figure}

This section presents \system{}'s request planner. Minimizing $W$ and
minimizing $B$ are the two endpoints of the same planning problem, and
each admits an exact algorithm. We solve both, then run Adaptive
Frontier Search between them, and finally relate the resulting merge
plans to object-store charges. As shown in\figref{fig:planning-frontier},
the feasible merge plans form a point cloud in the $(B,W)$ plane, and
for finite $P$ an optimal plan is attained at the lower convex hull
vertex with the smallest $T_P$.

\subsection{Minimizing \texorpdfstring{$W$}{W}}
\label{sec:exact-serial}

At the serial endpoint, requests do not overlap, so the planning
objective reduces to $W(\pi)$. The problem is therefore
\begin{equation*}
\pi^{*}_{\mathrm{serial}}
\in \argmin_{\pi} W(\pi)
\quad
\text{s.t.}\quad
K(\pi) \le K_{\max}.
\end{equation*}

\subsubsection{Basic Dynamic Program}
We begin with the exact Sum-Minimizing DP. It reduces to a classic partition problem: cut $n$
sorted intervals into at most $K_{\max}$ contiguous groups and minimize
the sum of request-time estimates. Let $c(t+1,i) = \tau(r_i - l_{t+1})$
denote the predicted time of merging required intervals $t+1,\dots,i$ into
one request. Define $dp[i][k]$ as the minimum sum of estimates to cover
the first $i$ intervals with exactly $k$ requests. The recurrence is
\begin{equation*}
dp[i][k] =
\begin{cases}
0, & \text{if } i = k = 0, \\[4pt]
\min\limits_{k-1 \le t < i}
\bigl\{
dp[t][k{-}1] + c(t{+}1,i)
\bigr\},
& \text{if } 1 \le k \le i, \\[4pt]
+\infty, & \text{otherwise}.
\end{cases}
\end{equation*}
The optimal sum is $\min_{1 \le k \le K_{\max}} dp[n][k]$, and
recording each state's minimizing split lets the plan be recovered by
backtracking. The DP has $n K_{\max}$
states, each with an $O(n)$ inner search over split points $t$, for a
total of $O(n^2 K_{\max})$ time.

\subsubsection{Divide-and-Conquer DP Optimization}
\layoutmargin{25.52394pt}{5.98949pt}{2.64598pt}
\papertext{In the same-region setting that is our primary focus,
\Cref{fig:bg-calibration} shows that a convex $\tau$ closely
approximates measured single-request latency. We therefore use the
convex fit in this setting and apply the divide-and-conquer optimization
described below.}
The inner search over the final split point is what makes the basic DP
quadratic in $n$. For a fixed request count $k$, let
$\mathrm{opt}[i][k]$ be the smallest split point attaining $dp[i][k]$.
Convexity of $\tau$ makes $c(a,b)=\tau(r_b-l_a)$ satisfy the quadrangle
inequality~\cite{quad_rangleyao1980efficient,quad_WACHS1989518},
\begin{equation}
c(a,c) + c(b,d) \le c(a,d) + c(b,c)
\quad \text{for all } a \le b \le c \le d,
\label{eq:quadrangle}
\end{equation}
which implies
$\mathrm{opt}[i][k] \le \mathrm{opt}[i{+}1][k]$. Divide-and-conquer DP
optimization exploits this monotonicity. After finding the optimal split
$t^*$ for a midpoint row, the recursion searches only split points at most
$t^*$ in the left half and at least $t^*$ in the right half. \papertext{For a
nondecreasing but nonconvex surrogate, the basic DP remains exact but
cannot use this acceleration.}

\textit{Thresholded variant.} Finite-$P$ planning in\secref{sec:finite}
repeatedly solves a constrained version of this DP. Given a threshold
$\beta$, it minimizes $W(\pi)$ subject to $B(\pi) \le \beta$. Equivalently,
every planned range must have request-time estimate at most $\beta$, so a
transition ending at interval $i$ after split $t$ is feasible only when
$c(t{+}1,i) \le \beta$. Setting $\beta=\infty$ recovers the unconstrained
Sum-Minimizing DP.

The threshold restricts the feasible split points but preserves their
monotone structure. For each endpoint $i$, let $t_\beta(i)$ be the earliest
split that makes the final request feasible under threshold $\beta$,
\begin{equation*}
t_\beta(i)=\min\{t:c(t{+}1,i)\le\beta\}.
\end{equation*}
The feasible splits are therefore $t_\beta(i)\le t<i$. Because extending the
covered prefix cannot move this boundary to the left, $t_\beta(i)$ is
nondecreasing in $i$. Together with the quadrangle inequality,
this implies that the smallest optimal feasible split remains nondecreasing.
Therefore, the thresholded DP can use the same divide-and-conquer
optimization, with the candidate splits for row $i$ additionally restricted
to $t \ge t_\beta(i)$.

The threshold can also make some DP rows infeasible. The feasible rows in
layer $k$ form an initial contiguous range $[k,h_k]$, where $h_k$ is the
largest interval index reachable by a feasible exactly-$k$ plan. Starting
from $h_0=0$, we extend
each $h_k$ from $h_{k-1}$ while the final request remains under $\beta$, and
keep it at $n$ once all intervals are covered. For each midpoint row, we
further bound the candidate splits by $t_\beta(i)$, $h_{k-1}$, and the
bounds inherited from the divide-and-conquer recursion.
\Cref{alg:dc_dp} gives the procedure.

\begin{algorithm}[t]
\caption{Exact Sum-Minimizing DP with D\&C optimization}
\label{alg:dc_dp}
\begin{algorithmic}[1]
\Require Sorted intervals $\mathcal{I}=\{[l_i,r_i)\}_{i=1}^{n}$
\Require Request budget $K_{\max}$ and feasible cap $\beta$
\Ensure Minimum $W$ and the corresponding merge plan
\Function{ComputeLayer}{$L$, $R$, $optL$, $optR$, $k$}
  \If{$L > R$} \Return \EndIf
  \State $mid \gets \lfloor (L + R) / 2 \rfloor$
  \State $lo \gets \max(optL,\, k{-}1,\, t_\beta[mid])$
  \State $hi \gets \min(optR,\, mid{-}1,\, h[k{-}1])$
  \State $dp[mid][k] \gets +\infty$
  \State $t^{*} \gets lo$
  \For{$t \gets lo$ to $hi$}
    \State $\mathrm{cand} \gets dp[t][k{-}1] + c(t{+}1, mid)$
    \If{$\mathrm{cand} < dp[mid][k]$}
      \State $dp[mid][k] \gets \mathrm{cand}$
      \State $t^{*} \gets t$
    \EndIf
  \EndFor
  \State $\mathrm{prev}[mid][k] \gets t^{*}$
  \State \Call{ComputeLayer}{$L$, $mid{-}1$, $optL$, $t^{*}$, $k$}
  \State \Call{ComputeLayer}{$mid{+}1$, $R$, $t^{*}$, $optR$, $k$}
\EndFunction
\Statex
\Function{DCSumDP}{$\mathcal{I}, K_{\max}, \beta$}
  \State $n \gets |\mathcal{I}|$
  \State $K_{\max} \gets \min(K_{\max}, n)$
  \State Initialize $dp$ to $+\infty$ and $\mathrm{prev}$ to $-1$
  \State $dp[0][0] \gets 0$
  \For{$i \gets 1$ to $n$}
    \State $t_\beta[i] \gets \min\{0\le t<i:c(t{+}1,i)\le\beta\}$
  \EndFor
  \State $h[0] \gets 0$
  \For{$k \gets 1$ to $K_{\max}$}
    \State $h[k] \gets h[k{-}1]$
    \State $s \gets h[k{-}1]+1$
    \While{$h[k]<n$ \textbf{and} $c(s,h[k]{+}1)\le\beta$}
      \State $h[k] \gets h[k]+1$
    \EndWhile
    \State \Call{ComputeLayer}{$k$, $h[k]$, $k{-}1$, $h[k{-}1]$, $k$}
  \EndFor
  \State $k^{*} \gets \argmin_{1 \le k \le K_{\max}} dp[n][k]$
  \State $\pi^{*} \gets$ \Call{Backtrack}{$\mathrm{prev},n,k^{*}$}
  \State \Return $(dp[n][k^{*}], \pi^{*})$
\EndFunction
\end{algorithmic}
\vspace{-1pt}
\end{algorithm}

\textit{Complexity.} A two-pointer scan computes all $t_\beta(i)$ in $O(n)$
time. Since $h_k$ never decreases, all layer boundaries can also be found in
$O(n+K_{\max})$ time. Each layer's divide-and-conquer recursion visits
$O(h_k)\le O(n)$ rows and performs $O(h_k\log n)$ total candidate work.
With divide-and-conquer optimization, the Sum-Minimizing DP therefore runs in
$O(n K_{\max} \log n)$ time and $O(n K_{\max})$ space.

Stronger Monge-based techniques~\cite{monge_aggarwal1986geometric} can remove
the $\log n$ factor in the uncapped case, but we use the divide-and-conquer
form because it naturally extends to the bottleneck-constrained subproblems
in~\secref{sec:finite}.
Alternatively, one can employ a Lagrangian relaxation of the cardinality
constraint and binary-search on the associated penalty
parameter~\cite{alien_opt_swiss_wiki}, removing the linear dependence on
$K_{\max}$ and replacing it with a $\log\Lambda$ factor where $\Lambda$ is
the penalty search range. Finding the exact optimum requires a fine-grained
search over $\Lambda$, making the constant large in practice. However, when
an approximate solution suffices, the search can stop early, trading a
small loss in optimality for far fewer iterations. Because of the cost of
the fine-grained exact search, we keep D\&C as the default exact backend.
We use Lagrangian relaxation with early stopping only as a fast path
(``+Lagrangian'' in \figref{fig:eval-planner-overhead-breakdown}) in the
approximate online planner when $K_{\max}$ is large enough that D\&C's linear
dependence on $K_{\max}$ dominates the planner cost.

\subsection{Minimizing \texorpdfstring{$B$}{B}}
\label{sec:exact-bottleneck}

Under ideal full parallelism, query completion time is determined by
the slowest request, so the objective is
\begin{equation*}
\pi^{*}_{\mathrm{parallel}}
\in \argmin_{\pi: K(\pi) \le K_{\max}} B(\pi).
\end{equation*}
Since $\tau(\ell)$ is nondecreasing, minimizing the maximum request-time estimate
reduces to minimizing the maximum merged span: we want the smallest
$\lambda$ such that the intervals admit a partition into at most
$K_{\max}$ groups each with span $\le \lambda$.

We compute this $\lambda$ by binary search. For a candidate $\lambda$,
a greedy left-to-right scan tests feasibility: it extends the current
group with successive intervals while the group's span stays
$\le \lambda$, and starts a new group whenever extending would exceed
$\lambda$. $\lambda$ is feasible exactly when the scan produces at most $K_{\max}$
groups. Feasibility is monotone in $\lambda$ (a larger $\lambda$
admits any plan a smaller one does), so binary search converges on
the smallest feasible $\lambda$. \cref{alg:parallel} shows the full
procedure.

\textit{Complexity.} Each feasibility test is $O(n)$, and the value-domain binary search takes
$O(n \log U)$ time, where $U = r_n - l_1$ is the search range of merged spans.

\begin{algorithm}[t]
\caption{Exact $B$ minimization by span-threshold search}
\label{alg:parallel}
\begin{algorithmic}[1]
\Require Sorted intervals $[l_i,r_i)$, budget $K_{\max}$
\Ensure $\min B$ objective value and corresponding merge plan
\State $L \gets \max_i(r_i - l_i)$,\quad $R \gets r_n - l_1$
\While{$L < R$}
  \State $\mathit{mid} \gets \lfloor (L + R) / 2 \rfloor$
  \If{\Call{Feasible}{$\mathit{mid}$}} $R \gets \mathit{mid}$
  \Else\ $L \gets \mathit{mid} + 1$ \EndIf
\EndWhile
\State $\pi^{*} \gets$ greedy partition under span cap $L$
\State \Return $(B(\pi^{*}), \pi^{*})$
\Statex
\Function{Feasible}{$\lambda$}
  \State $k \gets 0$,\quad $s \gets 1$
  \While{$s \le n$}
    \State $k \gets k + 1$,\quad $i \gets s$
    \While{$i \le n$ \textbf{and} $r_i - l_s \le \lambda$} $i \gets i + 1$ \EndWhile
    \State $s \gets i$
    \If{$k > K_{\max}$} \Return false \EndIf
  \EndWhile
  \State \Return true
\EndFunction
\end{algorithmic}
\end{algorithm}

\subsection{Finite \texorpdfstring{$P$}{P}: Searching the Frontier}
\label{sec:finite}

We now consider the case of bounded parallelism. The objective
$T_P(\pi)$ defined in\secref{sec:formulation} combines $W$ and
$B$, so neither the $\min W$ planner from\secref{sec:exact-serial} nor
the $\min B$ planner from\secref{sec:exact-bottleneck} is optimal on
its own. We treat the bottleneck cap $\beta$ as a search parameter and
define
\begin{equation*}
W^{*}(\beta)
=
\min_{\pi: K(\pi) \le K_{\max},\, B(\pi) \le \beta}
W(\pi),
\end{equation*}
which the thresholded Sum-Minimizing DP from\secref{sec:exact-serial}
evaluates for a single $\beta$. As $\beta$ increases, the feasible set
expands and $W^{*}(\beta)$ is nonincreasing, so the evaluated plans
trace a $(B, W)$ frontier. Its two ends are the $\min B$ plan
and the $\min W$ plan ($\beta = \infty$).
The planner searches this frontier for the plan minimizing $T_P$.

An exact solver could enumerate every distinct candidate $\beta$.
There are $O(n^2)$ distinct segment costs in the worst case, and each
thresholded DP evaluation costs $O(n K_{\max}\log n)$, so enumeration
takes $O(n^3 K_{\max}\log n)$. This is exact but too expensive online,
which motivates the $(1{+}\epsilon)$-approximate adaptive search below.

\subsubsection{Adaptive Frontier Search}
\label{sec:finite-adaptive}

$T_P$ is linear with nonnegative coefficients on both $B$ and $W$, so
its minimum over the feasible $(B, W)$ point cloud is attained at a
vertex of the cloud's lower convex hull, the convex lower-left
boundary of the cloud.
The hull can still contain many points, so we allow a
$(1{+}\epsilon)$-approximation: the search stops once no unexplored
frontier segment can improve the current best by more than a factor
of $1{+}\epsilon$. Smaller $\epsilon$ tightens the stopping
criterion, visiting more frontier points for a sharper bound on
$T_P$. Larger $\epsilon$ stops sooner with a coarser bound.

The search starts from the two extreme plans:
\cref{alg:parallel} gives the $\min B$ endpoint and \cref{alg:dc_dp}
with $\beta=\infty$ gives the $\min W$ endpoint. These two points
define the initial active segment. More generally, an active segment is
represented by a left cap $\beta_\ell$ with $W_\ell=W^*(\beta_\ell)$ and an
evaluated plan $p_r=(B_r,W_r)$ as its local right endpoint. We denote the left
cap state by $q_\ell=(\beta_\ell,W_\ell)$, although this pair need not
correspond to an evaluated plan point. Any undiscovered plan in the segment
has $B\ge\beta_\ell$ and $W\ge W_r$.
For a point $p=(B,W)$, we write $T_P(p)=T_P(B,W)$ for the corresponding value
of the planning objective.
Since $T_P$ is nondecreasing in both arguments,
\begin{equation*}
\operatorname{LB}(q_\ell,p_r) := T_P(\beta_\ell,W_r)
\end{equation*}
lower-bounds every unexplored plan in that segment.

The algorithm expands the segment with the smallest lower bound and evaluates
the midpoint cap $\beta=(\beta_\ell+B_r)/2$ with the thresholded DP. The DP
returns a witness plan $\pi_m$ with actual point
$p_m=(B(\pi_m),W_m)$, where $B(\pi_m)\le\beta$ and
$W_m=W^*(\beta)$. If $B(\pi_m)>\beta_\ell$, this is a new plan point inside
the segment, so the algorithm updates the current best and splits the segment
at $p_m$. Otherwise, $\pi_m$ was already feasible at the left cap, which
implies $W_m=W_\ell$. No frontier improvement exists on
$(\beta_\ell,\beta]$, so the algorithm drops this interval and advances the
left cap state to $(\beta,W_\ell)$ while retaining the same local right
endpoint. \Cref{alg:finite} gives the procedure.

\begin{algorithm}[t]
\caption{Adaptive Frontier Search for finite $P$}
\label{alg:finite}
\begin{algorithmic}[1]
\Require Intervals $\mathcal I$, budget $K_{\max}$, knob $P$, tolerance $\epsilon$
\Ensure $(1{+}\epsilon)$-approximate plan under $T_P$ upon termination
\State $(b_L, \_) \gets \Call{MinBPlan}{\mathcal I, K_{\max}}$ \Comment{$\min B$, \cref{alg:parallel}}
\State $(W_L, \pi_L) \gets \Call{DCSumDP}{\mathcal I, K_{\max}, b_L}$ \Comment{$\min W$ at $B{=}b_L$}
\State $p_L \gets (b_L, W_L)$
\State $(W_R, \pi_R) \gets \Call{DCSumDP}{\mathcal I, K_{\max}, \infty}$ \Comment{$\min W$, \cref{alg:dc_dp}}
\State $p_R \gets (B(\pi_R), W_R)$
\State $(z_{\mathrm{best}}, \pi_{\mathrm{best}}) \gets$ better of $(T_P(p_L), \pi_L)$ and $(T_P(p_R), \pi_R)$
\State $\mathcal H \gets$ min-heap keyed by $\operatorname{LB}$,\quad $\mathcal H.\mathrm{push}((p_L, p_R))$
\While{$\mathcal H$ not empty}
  \State $(q_\ell, p_r) \gets \mathcal H.\mathrm{pop\_min}()$
  \State $(\beta_\ell, W_\ell) \gets q_\ell$,\quad $(B_r, W_r) \gets p_r$
  \If{$z_{\mathrm{best}} \le (1{+}\epsilon)\,\operatorname{LB}(q_\ell, p_r)$}
    \State \Return $(z_{\mathrm{best}}, \pi_{\mathrm{best}})$
  \EndIf
  \State $\beta \gets (\beta_\ell + B_r)/2$
  \State $(W_m, \pi_m) \gets \Call{DCSumDP}{\mathcal I, K_{\max}, \beta}$
  \If{$B(\pi_m) > \beta_\ell$} \Comment{split at the new plan point}
    \State $p_m \gets (B(\pi_m), W_m)$
    \State $z_m \gets T_P(p_m)$
    \If{$z_m < z_{\mathrm{best}}$}
      \State $z_{\mathrm{best}} \gets z_m$
      \State $\pi_{\mathrm{best}} \gets \pi_m$
    \EndIf
    \State $\mathcal H.\mathrm{push}((q_\ell, p_m))$
    \State $\mathcal H.\mathrm{push}((p_m, p_r))$
  \Else \Comment{$W^*$ flat on $(\beta_\ell, \beta]$: advance left cap}
    \State $\mathcal H.\mathrm{push}(((\beta, W_\ell),\, p_r))$
  \EndIf
\EndWhile
\State \Return $(z_{\mathrm{best}}, \pi_{\mathrm{best}})$
\end{algorithmic}
\end{algorithm}

\subsubsection{Complexity and Guarantees}
\label{sec:planning-complexity}

Each frontier expansion evaluates one thresholded Sum-Minimizing DP. The
best-first search therefore costs $O(TnK_{\max}\log n)$, where $T$ is the
number of expansions. Each expansion either discovers a new frontier point or
halves the unexplored $B$ range of one segment. Only finitely many frontier
points exist, and repeated drops make the lower bound converge geometrically
to the objective of a feasible right-endpoint plan, which the current best
cannot exceed. The search therefore terminates for $\epsilon>0$. Because the
heap always expands the segment with the smallest lower bound, the stopping
test also guarantees that the returned plan is within a factor
$1{+}\epsilon$ of the optimum. Empirically, best-first stops after a few
expansions, and with a few hundred intervals after probing and moderate
$K_{\max}$, the planner finishes in single-digit milliseconds per query.

\textit{Worst-case-bounded variant.} Let $B_{\min}>0$ be the minimum feasible
$B$, and let $B_{\max}$ be the $B$ value of the $\min W$ plan. These values are
given by the two endpoint plans used to initialize best-first search. The
bounded variant evaluates the thresholded DP at caps
$\beta_j=\min\{B_{\max},(1{+}\epsilon)^jB_{\min}\}$, starting from $j=0$ and
stopping when the cap reaches $B_{\max}$. It returns the evaluated plan with
the smallest $T_P$. For an optimal point $p^*=(B^*,W^*)$, let $\beta_j$ be the
first cap no smaller than $B^*$. Then $\beta_j\le(1{+}\epsilon)B^*$, and because
$p^*$ is feasible under $\beta_j$, the DP returns a plan
$p_j=(B_j,W_j)$ with $B_j\le\beta_j$ and $W_j\le W^*$. The nonnegative
coefficients of $T_P$ then give
$T_P(p_j)\le(1{+}\epsilon)T_P(p^*)$. The grid contains at most
$1+\lceil\log_{1+\epsilon}(B_{\max}/B_{\min})\rceil$ caps. Its total running
time is
\begin{equation*}
O\!\left(
  nK_{\max}\log n
  \cdot
  \left[
    1 + \frac{\log(B_{\max}/B_{\min})}{\log(1+\epsilon)}
  \right]
\right).
\end{equation*}

\subsection{Cost Accounting}
\label{sec:planning-cost}

\papertext{A merge plan determines both the number of payload GETs and the
amount of data read.\layoutmargin{27.72894pt}{5.98949pt}{2.64598pt} Major cloud object stores,
such as Amazon S3,
Google Cloud Storage, and Azure Blob Storage, charge for GET requests and
may also charge for cross-region data
transfer~\cite{amazonS3Pricing,googleCloudStorage,azureBlobStorage}. These
services waive data-transfer fees within the same region. Thus, in a
same-region deployment, the cost is the number of payload GETs multiplied
by the price per GET. In a cross-region deployment, the cost also includes
the number of bytes read multiplied by the data-transfer price per byte for
that deployment. We focus on the common same-region setting.}

\papertext{We formalize these two cases as follows. Let $p_{\mathrm{GET}}$ be the price per payload GET and
$p_{\mathrm{byte}}$ be the data-transfer price per byte for that deployment.
A merge plan $\pi$ issues $K(\pi)$ payload GETs, where $\ell_u$ is the
length of its $u$th planned range. Then
\begin{equation*}
\operatorname{Cost}(\pi)=
\begin{cases}
p_{\mathrm{GET}}K(\pi),
& \text{same-region}, \\
p_{\mathrm{GET}}K(\pi)
+p_{\mathrm{byte}}\displaystyle\sum_{u=1}^{K(\pi)}\ell_u,
& \text{cross-region}.
\end{cases}
\end{equation*}
In our end-to-end experiments, we report the cost under both pricing cases in
\Cref{sec:eval-end-to-end-performance}.}

\section{Cluster Payload Reordering}
\label{sec:reordering}

\begin{figure}[!t]
  \centering
  \includegraphics[width=0.92\linewidth]{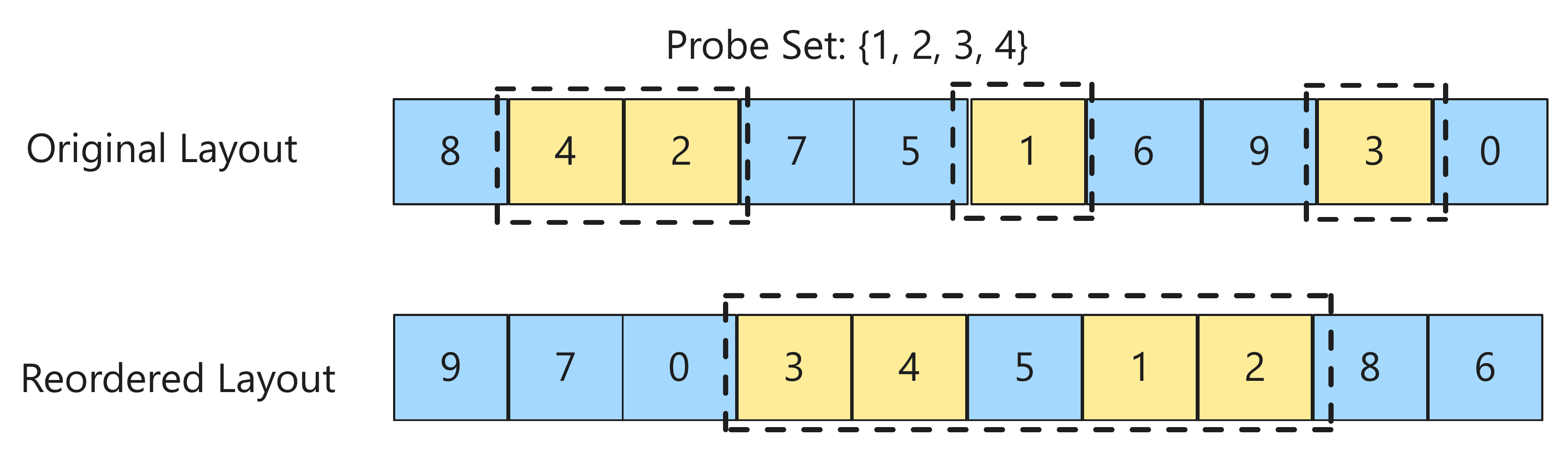}
  \caption{\textbf{Effect of reordering.} Frequently co-accessed clusters are
  placed closer together, reducing the gaps between required intervals;
  each box denotes one cluster in contiguous layout.}
  \Description{A two-row schematic: top row shows the original layout with
  large gaps between co-accessed clusters. The bottom row shows the reordered
  layout with those clusters placed closer together.}
  \label{fig:reorder-schematic}
\end{figure}

\layoutmargin{27.72894pt}{5.98949pt}{2.64598pt}
\papertext{Request planning decides how to group required byte intervals into
requests, but it does not change the cluster layout. By exploiting recurring
cluster co-access patterns, cluster payload reordering places frequently
co-accessed clusters closer together, thereby reducing the gaps between
required intervals. We call this property co-access locality.
In \figref{fig:reorder-schematic}, the original layout needs three requests to
fetch the required intervals without reading the large gaps between them. After
reordering, one request can fetch all required intervals while overfetching only
one extra cluster payload.}

\begin{figure*}[!tbp]
  \centering
  \layoutmarginfloat{27.72894pt}{24.22044pt}{2.64598pt}%
  \includegraphics[width=\linewidth]{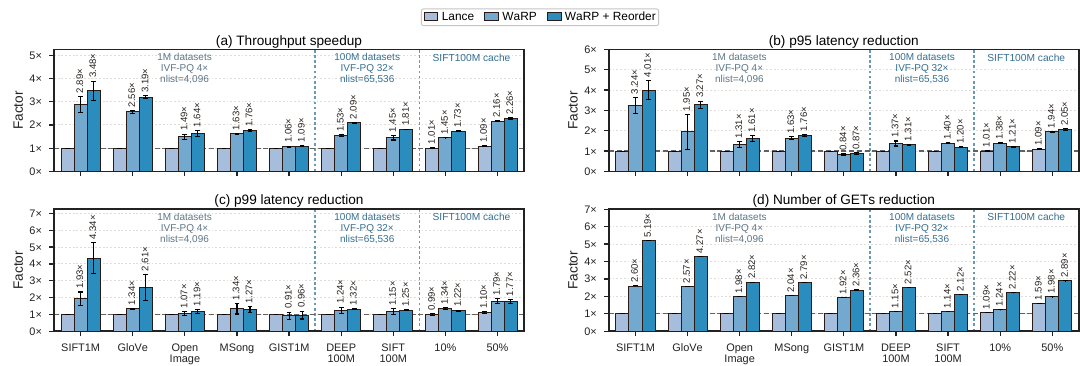}
  \caption{\papertext{\textbf{Normalized end-to-end performance.}
  Results for each dataset are normalized to Lance. The
  10\% and 50\% SIFT100M cache results instead use cache-off SIFT100M Lance as
  the reference. Error bars show propagated sample standard
  deviations across three runs.}}
  \Description{Four bar charts compare Lance, WaRP, and WaRP plus Reorder on
  seven datasets and two cache budgets. They show higher-is-better throughput,
  p95/p99 latency, and GET reduction. WaRP variants usually improve all metrics
  except GIST1M latency.}
  \label{fig:eval-real-lance-metrics}
\end{figure*}

\begin{figure*}[!t]
  \centering
  \layoutmarginfloat{27.72894pt}{15.10497pt}{2.64598pt}%
  \includegraphics[width=\textwidth]{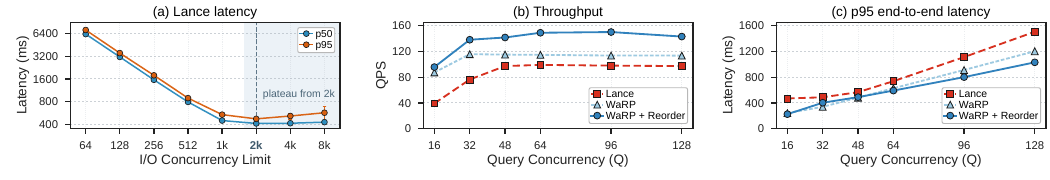}
  \caption{\papertext{\textbf{SIFT100M concurrency scaling.}
  Panel (a) varies the I/O concurrency limit for Lance, while panels
  (b) and (c) vary query concurrency for all three methods.}}
  \Description{Three SIFT100M line charts. The left panel plots Lance p50 and
  p95 latency as the I/O concurrency limit increases from 64 to 8,192 at
  query concurrency 32. The middle and right panels compare Lance, WaRP, and
  WaRP plus Reorder as query concurrency increases from 16 to 128 at an I/O
  concurrency limit of 2,048. They report QPS and p95 end-to-end latency.}
  \label{fig:eval-real-lance-concurrency-scaling}
\end{figure*}

\subsection{Layout Materialization and Maintenance}
\label{sec:reorder-materialization}

\layoutmargin{27.72894pt}{5.98949pt}{2.64598pt}
\papertext{Cluster reordering runs offline during index build or rewrite.
Throughout our evaluation, we compute one cluster permutation for the entire
IVF index. We materialize the resulting cluster layout by emitting payloads in
that order during the existing payload-writing phase. Each payload is still
written once, so reordering requires no additional rewrite or upload pass. Its
only added work is computing the permutation, which
\Cref{sec:eval-reorder} shows takes milliseconds for thousands of clusters.}

\papertext{For indexes with incremental updates, newly added data gradually weakens
co-access locality. Such indexes typically use rebuilds or segment merges to
maintain index quality. During these operations, the affected index data is
rewritten. We recompute the cluster permutation and apply it when writing this
data. Thus, we update the reordered cluster layout as part of normal index
maintenance without a separate rewrite pass. The next two subsections explain
how we measure co-access locality and compute the permutation.}

\subsection{Measuring Co-Access Locality}
\label{sec:reorder-objective}

\layoutmargin{27.72894pt}{5.98949pt}{2.64598pt}
\papertext{
Clusters that often appear in the same probe set should be close in the layout.
To compare cluster layouts, we use a local query set to measure how often each
cluster pair is selected together. } We combine this frequency with the byte gap
between the pair in each layout to measure co-access locality. Let $\sigma$ be a
permutation of the $C$ clusters, and let $\mathcal{Q}_{\mathrm{layout}}$ be the
local query set. For clusters $i$ and $j$, we define their \emph{co-access
weight} as
\begin{equation}
a_{ij} = \frac{1}{|\mathcal{Q}_{\mathrm{layout}}|}
\sum_{q \in \mathcal{Q}_{\mathrm{layout}}}
  \mathbf{1}[i \in S_q \wedge j \in S_q],
\label{eq:coaccess-weight}
\end{equation}
where $S_q$ is the probe set of query $q$.

For a permutation $\sigma$, we define the \emph{weighted co-access gap} as
\begin{equation*}
G(\sigma) = \sum_{i < j} a_{ij} \cdot \mathrm{gap}_{\sigma}(i,j),
\end{equation*}
where $\mathrm{gap}_{\sigma}(i,j)=|f_{\sigma}(i)-f_{\sigma}(j)|$ and
$f_{\sigma}(i)$ is the byte offset of cluster $i$.
A smaller $G(\sigma)$ means that frequently co-accessed clusters are placed
closer together. This metric is independent of how $\sigma$ is constructed,
allowing us to compare layouts derived from centroid geometry or query logs.
Directly minimizing $G(\sigma)$ is an instance of the \emph{minimum linear
arrangement} problem~\cite{hilbert1,garey1974some} and is NP-hard, so we study
practical reordering methods.

\subsection{Reordering Methods}
\label{sec:reorder-methods}
\layoutmargin{27.72894pt}{5.98949pt}{2.64598pt}
\papertext{
  We compute a cluster permutation before materializing the cluster layout. Each
method uses either the cluster centroids already available for query routing or
co-access counts from a query log to place frequently co-accessed clusters
nearby in the layout. }
We compare random projection, principal component
analysis (PCA), space-filling curves (SFCs), and spectral ordering.
\Cref{tab:reorder-methods} summarizes their computation costs for $C$
centroids of dimension $d$.

\begin{table}[H]
\caption{\textbf{Reordering Methods.}}
\label{tab:reorder-methods}
\vspace{-4pt}
\centering
\small
\begin{tabular*}{\linewidth}{@{\extracolsep{\fill}}lll}
\toprule
Method & Time & Description \\
\midrule
Random projection       & $O(Cd)$   & Sort by random projection \\
PCA-1D                  & $O(Cd^2)$ & Sort by first PCA axis \\
PCA $n$-D + SFC         & $O(Cd^2)$ & Order in top-$n$ PCA space \\
Spectral (cold)         & $O(C^3)$  & Spectral order from centroid similarity \\
Spectral (log)          & $O(C^3)$  & Spectral order from co-probe counts \\
\bottomrule
\end{tabular*}
\end{table}

Random projection orders clusters by their projected values along a random
direction. PCA-1D~\cite{PCA} instead uses the first principal component of the
cluster centroids. PCA $n$-D + SFC projects the centroids onto their top-$n$
components and orders the projected coordinates with either Z-order
(Morton)~\cite{zordermorton1966computer} or a Hilbert
curve~\cite{kamel1994hilbert}.
Spectral (cold) constructs a centroid-similarity graph and orders clusters by
its Laplacian's second eigenvector~\cite{Spectralfiedler1973algebraic}.
Spectral (log) applies the same ordering to a graph weighted by empirical
co-probe counts from $\mathcal{Q}_{\mathrm{layout}}$.

We use PCA $n$-D + Z-order by default because it captures multidimensional
centroid locality, needs no query log, and avoids the spectral methods' cubic
cost. Using more PCA components retains more information, but makes it harder
for a one-dimensional SFC order to keep nearby clusters together. We choose
$n$ to balance these two effects. Systems with reliable query logs and
tolerance for higher offline cost can instead use Spectral (log) to incorporate
workload information.

\section{End-to-End Evaluation}
\label{sec:evaluation}
\label{sec:eval-real-lance}

\begin{table}[!tbp]
\caption{\textbf{Evaluation Datasets and Workloads.}}
\label{tab:eval-datasets}
\vspace{-4pt}
\centering
\small
\begin{tabular*}{\linewidth}{@{\extracolsep{\fill}}lcccc}
\toprule
Name & Domain & Dim. & Vectors & Raw Size \\
\midrule
SIFT1M~\cite{ivfPQsift} & Image & 128 & 1M & 488\,MB \\
SIFT100M~\cite{ivfPQsift} & Image & 128 & 100M & 51\,GB \\
GloVe~\cite{glove} & Text & 100 & 1.18M & 450\,MB \\
DEEP1M~\cite{babenko2016efficientdeep} & Image & 96 & 1M & 366\,MB \\
DEEP100M~\cite{babenko2016efficientdeep} & Image & 96 & 100M & 38\,GB \\
OpenImage~\cite{openimage} & Image & 512 & 1M & 1.9\,GB \\
MSong~\cite{MsongDataset} & Audio & 420 & 0.99M & 1.6\,GB \\
GIST1M~\cite{ivfPQsift} & Image & 960 & 1M & 3.6\,GB \\
Random1M & Synthetic & 128 & 1M & 488\,MB \\
\bottomrule
\end{tabular*}
\end{table}

\begin{table*}[!t]
\layoutmarginfloat{27.72894pt}{15.10497pt}{2.64598pt}%
\caption{\papertext{\textbf{End-to-end bytes, bandwidth, overfetch, and cost.}
Each entry reports Lance / \system{} / \system{} + Reorder.}}
\label{tab:eval-payload-accounting}
\centering
\small
\renewcommand{\arraystretch}{1.0}
\setlength{\tabcolsep}{1.5pt}
\begin{tabular*}{\textwidth}{@{\extracolsep{\fill}}lccccccc@{}}
\toprule
Metric & SIFT1M & GloVe & OpenImage & MSong & GIST1M & DEEP100M & SIFT100M \\
\midrule
Fetched (MiB/q) & 6.5/14.1/11.2 & 5.4/15.3/10.7 & 21.5/23.8/22.5 & 15.3/18.4/16.9 & 39.5/41.4/40.3 & 9.7/25.6/26.7 & 10.4/25.1/33.5 \\
Overfetch ($\times$) & 0.0/1.2/0.7 & 0.0/1.9/1.0 & 0.0/0.1/0.0 & 0.0/0.2/0.1 & 0.0/0.0/0.0 & 0.0/1.6/1.8 & 0.0/1.4/2.2 \\
Payload bandwidth (Gbps) & 5.4/34.4/32.7 & 5.0/36.9/32.1 & 18.3/30.2/31.4 & 14.6/28.6/28.4 & 27.8/30.9/31.0 & 6.2/25.2/36.0 & 6.7/23.4/38.8 \\
Same-region (cents) & 97.5/37.5/18.8 & 98.2/38.3/23.0 & 98.0/49.5/34.8 & 98.1/48.1/35.2 & 97.7/50.8/41.3 & 222.7/194.1/88.3 & 222.2/195.7/104.7 \\
Cross-region (\$) & 6.7/12.8/10.0 & 5.7/13.9/9.7 & 19.8/21.4/20.1 & 14.4/16.7/15.2 & 35.7/36.9/35.8 & 10.8/24.4/24.4 & 11.4/24.0/30.5 \\
\bottomrule
\end{tabular*}
\end{table*}

In this section, we integrate \system{} into Lance's I/O scheduler
(v2.1.0-beta.0)~\cite{lance} and evaluate end-to-end on AWS S3.
Production object-storage vector search systems such as
turbopuffer~\cite{turbopuffer2026architecture} and Amazon S3
Vectors~\cite{amazonS3Vectors} are closed services, so we cannot
instrument or modify them. Lance is the open-source system that comes
closest. It includes a native IVF reader and a concurrent Range-GET
scheduler over S3, and is in production use for vector search today.
\layoutmargin{25.52394pt}{5.98949pt}{2.64598pt}
\papertext{Lance can merge adjacent ranges within a batched I/O submission using a
gap threshold and a request-size cap. Its IVF reader, however, submits
probed clusters separately, so these heuristics do not apply across clusters.
We preserve this behavior in the end-to-end comparison.}

We first evaluate end-to-end performance in
\Cref{sec:eval-end-to-end-performance}. We then study the effect of bounded
caching in \Cref{sec:eval-caching} and concurrency scaling in
\Cref{sec:eval-concurrency-scaling}. We use microbenchmarks in
\Cref{sec:microbenchmarks} to examine request planning and reordering in more
detail.

\subsection{Experimental Setup}
\label{sec:eval-setup}

\textit{Datasets.} \Cref{tab:eval-datasets} lists nine datasets.
Eight are real-world benchmarks spanning image, text, and audio embeddings
from 96 to 960 dimensions. Random1M is a synthetic stress case comprising
one million i.i.d.\ Gaussian vectors with weak co-access locality.
\layoutmargin{25.52394pt}{5.98949pt}{2.64598pt}
\papertext{We evaluate end-to-end performance on both 100M-scale datasets and
five real 1M-scale datasets. To keep the broader microbenchmark suite
tractable, we run it on six 1M-scale datasets.}

\textit{System and measurement.} All experiments run on an AWS EC2
\texttt{c7gn.8xlarge} instance (32 vCPUs, 64\,GB RAM, and 100\,Gbps
network bandwidth)
in the Tokyo region, with an S3~\cite{amazonS3} bucket in the same region.
Lance's concurrent Range-GET implementation~\cite{lance} handles all
object-store I/O.
All runs use a closed-loop workload with two independent concurrency
controls. Query concurrency $Q$ sets the maximum number of active end-to-end
queries, controlling the query load. The I/O concurrency limit independently
caps the number of scheduled Range-GETs kept in flight, controlling I/O
parallelism. For the end-to-end experiments, unless otherwise noted, all
three methods use $Q=32$ and an I/O concurrency limit of 2048. For both
\system{} variants, we set the request parallelism estimate $P$ to 64 and
leave the request budget unbounded.
\layoutmargin{27.72894pt}{15.10497pt}{2.64598pt}
\papertext{The five 1M-scale datasets use 4$\times$-compressed IVFPQ with
4,096 partitions and $nprobe=128$. SIFT100M and DEEP100M use
32$\times$-compressed IVFPQ with 65,536 partitions, $nprobe=256$, and
refinement factor 8.} For \system{} + Reorder, we use PCA $n$-D + Z-order
and choose the smallest $n \le 8$ whose top-$n$ components explain at least
$35\%$ of centroid variance.
\layoutmargin{27.72894pt}{5.98949pt}{2.64598pt}
\papertext{For each dataset, we run each method three times in counterbalanced
order, using 200 warm-up queries followed by 10,000 measured queries per run.}

\subsection{Overall Performance}
\label{sec:eval-end-to-end-performance}

Across seven real-world datasets, we measure the end-to-end impact of request
planning and reordering.
\layoutmargin{25.52394pt}{5.98949pt}{2.64598pt}
\papertext{\Cref{fig:eval-real-lance-concurrency-scaling}a shows that Lance's
p50 and p95 latencies plateau by an I/O concurrency limit of 2048, so this
setting does not limit performance.}

\layoutmargin{27.72894pt}{5.98949pt}{2.64598pt}
\papertext{Across all seven datasets in \Cref{fig:eval-real-lance-metrics},
both \system{} variants improve throughput and reduce GETs relative to
Lance. \system{} alone reaches $1.06\text{--}2.89\times$ Lance's QPS and
reduces GETs by $1.14\text{--}2.60\times$. Adding reordering widens these
ranges to $1.09\text{--}3.48\times$ and $2.12\text{--}5.19\times$.
Both variants also reduce p95 and p99 latency on six of the seven datasets.
The exception is GIST1M. Its 960-dimensional vectors make each cluster
substantially larger, so each GET already amortizes more fixed overhead and
fixed request overhead accounts for a smaller share of latency. A similar
effect appears for the high-dimensional datasets in
\Cref{sec:QPS-cost-eval}.}

\layoutmargin{27.72894pt}{15.10497pt}{2.64598pt}
\papertext{\Cref{tab:eval-payload-accounting} reports fetched bytes, overfetch,
payload bandwidth, and cost for the same 10,000-query workloads. Because
merging fetches extra bytes while increasing QPS, both \system{} variants
use more payload bandwidth than Lance. We apply Tokyo's S3 price of
\$0.0037 per 10,000 GETs. Same-region data transfer is free, while the
cross-region case adds \$0.09/GB for transfer to Northern
Virginia~\cite{amazonS3Pricing}. Both variants lower same-region cost on all
seven datasets, but their cross-region cost is higher than Lance's.}

\subsection{Effect of Caching}
\label{sec:eval-caching}

\layoutmargin{27.72894pt}{5.98949pt}{2.64598pt}
\papertext{We also evaluate request planning with a warm cluster cache.
The SIFT100M IVF index contains 10.3 GiB of cluster data, and we allocate
10\% and 50\% of this size to Lance's cluster cache. For each setting,
a fresh process first runs the same 10,000 queries to warm the cache and
then reports results from a second pass. In the three rightmost groups of
\Cref{fig:eval-real-lance-metrics}, a 10\% cache reduces remote GETs by
$4\text{--}8\%$, while throughput and p95 latency change little. At 50\%,
remote GETs fall by $27\text{--}43\%$, and all three methods improve in
throughput and p95 latency. Both \system{} variants still outperform Lance
at both cache sizes. Caching and request planning are therefore
complementary. Caching eliminates some remote reads, while request planning
reduces the cost of those that remain.}

\subsection{Concurrency Scaling}
\label{sec:eval-concurrency-scaling}

Finally, we test whether these gains persist as query load approaches
saturation.
\layoutmargin{27.72894pt}{15.10497pt}{2.64598pt}
\papertext{Using the SIFT100M setup from \Cref{sec:eval-end-to-end-performance}, we first vary Lance's I/O concurrency limit. As shown in \Cref{fig:eval-real-lance-concurrency-scaling}a, latency stops improving beyond roughly 2,000 concurrent requests. We therefore set the I/O concurrency limit to 2048 for the remaining experiments. \Cref{fig:eval-real-lance-concurrency-scaling}b and \Cref{fig:eval-real-lance-concurrency-scaling}c then vary query concurrency $Q$ from 16 to 128. Both \system{} variants achieve higher throughput and lower p95 latency than Lance across all tested concurrency levels. Lance peaks at about 99 QPS at $Q=64$, whereas \system{} + Reorder reaches about 150 QPS at $Q=96$. At $Q=128$, \system{} + Reorder still delivers $1.47\times$ Lance's QPS and reduces p95 latency by $1.46\times$, confirming that its advantage persists near saturation.}

\begin{figure*}[!tbp]
  \centering
  \includegraphics[width=\textwidth]{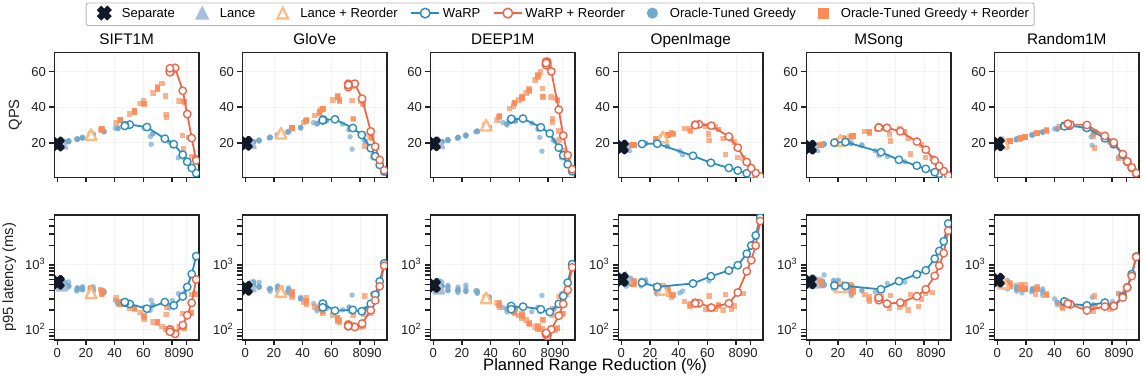}
  \caption{\textbf{I/O-plan throughput--reduction tradeoff.} QPS
  (top) and p95 latency (bottom) versus planned range reduction
  relative to Separate across six 1M-scale datasets
  ($nlist{=}4096$, $nprobe{=}128$, $Q{=}4$, I/O concurrency limit 64).}
  \Description{A twelve-panel figure arranged as a two-by-six grid over the same six datasets: SIFT1M, GloVe, DEEP1M, OpenImage, MSong, and Random1M. The top row shows achieved QPS and the bottom row shows p95 latency. Each panel contains \system{} and \system{} plus Reorder curves traced by varying K, Lance and Lance plus Reorder reference points, Separate, and Oracle-Tuned Greedy point clouds.}
  \label{fig:eval-qps-cost}
  \vspace{-1pt}
\end{figure*}

\section{Microbenchmarks}
\label{sec:microbenchmarks}
The remaining experiments isolate request planning and layout effects
from the rest of the ANN query pipeline.
\Cref{sec:QPS-cost-eval} examines the tradeoff between throughput and
planned range reduction.
\Cref{sec:eval-objective} tests robustness to planner parameters and
the deployability of greedy alternatives.
\Cref{sec:eval-quality} measures planner overhead and the contribution
of each optimization layer.
\Cref{sec:eval-reorder} compares reordering methods.
\Cref{sec:eval-scaling} tests sensitivity to probe count and cluster
granularity.

\textit{Setup.} Unless stated otherwise, the microbenchmarks use Faiss
(faiss-cpu v1.13.2)~\cite{faissivf} to build an IVF-Flat index and store
its cluster payloads in a single flat file on S3.
Within each experiment, all methods receive the same selected cluster
intervals and use the same I/O concurrency limit. Reordered variants
use PCA $n$-D + Z-order with $n$ selected as in
\Cref{sec:eval-end-to-end-performance}, except in
\Cref{sec:eval-reorder}, which compares the alternatives. We measure
I/O completion only, excluding distance computation and reranking.

\textit{Baselines.}
We compare \system{} with \textit{Separate}, \textit{Lance}, and
\textit{Oracle-Tuned Greedy}. \textit{Separate} issues one Range-GET per
probed cluster and serves as the reference for planned range reduction.
Many storage engines expose fixed thresholds for merging nearby reads.
\textit{Lance} uses its default settings, which merge
adjacent ranges when their gap is at most $64$\,KB and split merged
ranges larger than $16$\,MB. \layoutmargin{25.52394pt}{5.98949pt}{2.64598pt}
\papertext{\textit{Oracle-Tuned Greedy} tests a stronger fixed-threshold
policy. It sweeps gap thresholds from $64$\,KB to $16$\,MB and size
limits from $1$\,MB to $32$\,MB, doubling both at each step for 54
settings in total. It uses measured performance on each evaluation
workload to select the best setting, making it an optimistic reference
within the tested grid. We also pair Lance and Oracle-Tuned Greedy with
reordering to separate layout and request-planning effects.}

\subsection{Throughput and Planned Range Reduction}
\label{sec:QPS-cost-eval}

We then compare the methods at different planned range counts.
Across six datasets, we report how QPS and p95 latency change with
planned range reduction.

\papertext{All runs use query concurrency $Q{=}4$,\layoutmargin{25.52394pt}{5.98949pt}{2.64598pt}
an I/O concurrency limit
of 64, $P{=}16$, $nlist{=}4096$, and $nprobe{=}128$. To trace each
\system{} curve, we vary $K$ to obtain merge plans with different
planned range counts. Each Greedy point represents one gap/span setting,
while Lance and Separate each produce one point per dataset.}

\layoutmargin{25.52394pt}{5.98949pt}{2.64598pt}
\papertext{\Cref{fig:eval-qps-cost} shows that both too few and too many
planned ranges hurt performance. When the count is too low, the planner
merges distant intervals into large Range-GETs and therefore fetches more
unnecessary bytes.
Too many planned ranges create small Range-GETs that add per-request overhead
and consume I/O slots. An intermediate count maximizes QPS and minimizes p95
latency.}
Across all six datasets, \system{} + Reorder achieves the highest peak
QPS.
\layoutmargin{25.52394pt}{5.98949pt}{2.64598pt}
\papertext{Lance stays close to Separate under its default settings.
Reordering lifts it, but leaves a wide gap to \system{} + Reorder.
Oracle-Tuned Greedy matches \system{} + Reorder on several datasets
once tuned to its best setting. Its winning gap and span, however,
differ widely across datasets. The second experiment in
\Cref{sec:eval-objective} fixes one setting in advance and tests how
well it transfers.}

The combined benefit of request planning and reordering depends
strongly on co-access locality. It is largest on SIFT1M and DEEP1M,
where throughput reaches about $3\times$ that of Separate, and smaller
on GloVe. It is smallest on Random1M, whose i.i.d.\ Gaussian vectors
exhibit no co-access locality and leave reordering with nothing to
exploit.
\layoutmargin{25.52394pt}{5.98949pt}{2.64598pt}
\papertext{OpenImage and MSong fall below GloVe, reaching only
$1.6\text{--}1.7\times$ the QPS of Separate, but for a different reason. Under
IVF-Flat their 512- and 420-dimensional vectors make each cluster several
times larger, so every GET already amortizes
more of its fixed overhead and merging has less left to recover.
Product quantization could reduce per-cluster payload and increase the
benefit of merging.}\looseness=-1

\begin{figure*}[!tbp]
  \centering
  \layoutmarginfloat{25.52394pt}{5.98949pt}{2.64598pt}%
  \includegraphics[width=\textwidth]{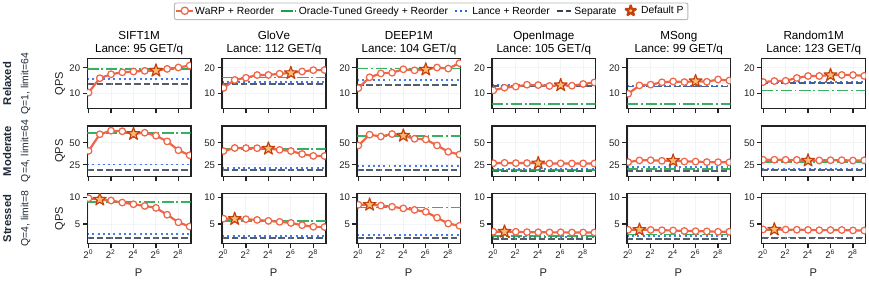}
  \caption{\papertext{\textbf{Closed-loop sensitivity to the Request
  Parallelism Estimate $P$.} \system{} + Reorder uses request budget
  $K_{\max}{=}64$. Oracle-Tuned Greedy + Reorder is restricted to
  settings with planned range count at most 64, whereas Lance + Reorder
  has no such restriction. Stars mark $P$ equal to the I/O concurrency
  limit divided by $Q$.}}
  \Description{A full-width figure arranged as a three-by-six grid of QPS plots. Rows are three runtime settings: relaxed with Q equal to 1 and I/O concurrency limit 64, moderate with Q equal to 4 and limit 64, and stressed with Q equal to 4 and limit 8. Columns are the six 1M-scale datasets: SIFT1M, GloVe, DEEP1M, OpenImage, MSong, and Random1M. Each column header annotates Lance's GETs per query. In every panel, WaRP plus Reorder uses request budget K max equal to 64. Oracle-Tuned Greedy plus Reorder is the best setting among those producing at most 64 planned ranges, while Lance plus Reorder is unbounded. WaRP is swept over P, and stars mark default P values of 64, 16, and 2.}
  \label{fig:eval-p-sweep-closedloop}
\end{figure*}

\subsection{Parameter Robustness}
\label{sec:eval-objective}
We first examine how the choice of $P$ affects \system{}
and whether a simple default works well across datasets and runtime
conditions. We then test whether Greedy's gap and span settings, once
tuned on one workload, remain effective when the query workload or
dataset changes.

As defined in \Cref{sec:formulation}, $P$ estimates how many requests
from one query can progress concurrently. We therefore use the I/O
concurrency limit divided by $Q$ as its default, corresponding to an
even share of the available I/O slots among concurrent queries.
Throughout the sweep, \system{} uses a request budget of
$K_{\max}{=}64$. We sweep $P$ under three runtime configurations.
The relaxed setting uses query concurrency $Q{=}1$ and an I/O
concurrency limit of 64, the moderate setting uses $Q{=}4$ and a limit
of 64, and the stressed setting uses $Q{=}4$ and a limit of 8.
\Cref{fig:eval-p-sweep-closedloop} shows the results.
QPS favors larger $P$ as more I/O concurrency becomes available to
each query. Stars mark the default $P$ in each runtime configuration.
Across the 18 panels, these
defaults achieve on average $97\%$ of the best QPS reached by any tested
$P$, so we can recommend this simple rule without per-workload tuning.

\begin{figure}[!tbp]
  \centering
  \layoutmarginfloat{25.52394pt}{5.98949pt}{2.64598pt}%
  \includegraphics[width=\linewidth]{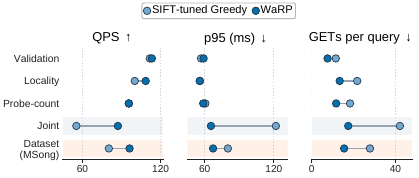}
  \caption{\papertext{\textbf{SIFT-Tuned Greedy under Shift.}
  The first four rows show the validation workload and three workload
  shifts on SIFT1M. The final row transfers both policies to MSong.
  Points are three-repeat medians.}}
  \Description{A single-column figure with three aligned dumbbell panels for QPS, p95 latency in milliseconds, and payload GETs per query. Light blue points show SIFT-tuned Greedy and dark blue points show WaRP. The five rows are the SIFT validation workload, a SIFT locality shift, a SIFT probe-count shift, a joint SIFT shift, and a dataset shift to MSong.}
  \label{fig:eval-greedy-locality-shift}
\end{figure}

\layoutmargin{25.52394pt}{5.98949pt}{2.64598pt}
\papertext{We next test whether a Greedy policy tuned on validation queries
still works when the workload changes. Using an IVFPQ index on
SIFT1M with 4,096 partitions, we select the best gap and span from the
same 54-setting grid used for Oracle-Tuned Greedy, but select them only
on the validation workload and then keep them fixed. We call the
resulting policy SIFT-tuned Greedy. We evaluate this fixed setting under
four shifts. In the locality shift, the same amount of required data is
spread over a byte span about six times wider. The probe-count shift
increases $nprobe$ from 32 to 256. The joint shift combines these two
changes, and the dataset shift applies the same setting to MSong. All
runs use $Q{=}4$ and an I/O concurrency limit of 64, with \system{} at
$P{=}32$ and $K_{\max}{=}128$.
\Cref{fig:eval-greedy-locality-shift} shows that SIFT-tuned Greedy
matches \system{} on the validation workload but falls behind \system{}
after the workload or dataset shifts.}

\subsection{Planner Overhead}
\label{sec:eval-quality}

\begin{figure}[!tbp]
  \centering
  \includegraphics[width=\linewidth]{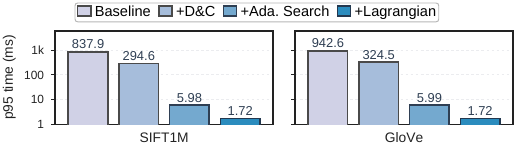}
  \caption{\textbf{Planner Overhead Breakdown.}
    Representative p95 planner time for $nprobe{=}256$ and $K_{\max}{=}128$.
  }
  \Description{A two-panel bar chart with a log-scale y-axis. Each panel shows four bars for the representative setting nprobe 256 and request budget K max 128. From left to right, the bars correspond to Baseline, then plus D and C, plus Ada Search, and plus Lagrangian, with each successive bar cumulatively adding one optimization stage over the previous one. Dataset names appear below each panel.}
  \label{fig:eval-planner-overhead-breakdown}
\end{figure}

\system{} plans on every query's critical path, so its own cost has to
stay well under the I/O it saves. A plain dynamic program does not,
because it enumerates candidate bottleneck values exhaustively and
takes hundreds of milliseconds. We therefore stack three optimization
layers, and \Cref{fig:eval-planner-overhead-breakdown} reports what
each one contributes at the representative $nprobe{=}256$, $K_{\max}{=}128$
setting on SIFT1M and GloVe. The unoptimized DP runs $838$\,ms on
SIFT1M and $943$\,ms on GloVe.
The divide-and-conquer backend (``+D\&C'') cuts those to $295$\,ms
and $325$\,ms, a $2.8\text{--}2.9\times$ saving on its own but still too slow for
the critical path. Adaptive frontier search (``+Ada.\ Search'') is
the largest speedup, dropping overhead to roughly $6$\,ms by
replacing exhaustive enumeration with a logarithmic-step search. The Lagrangian fast path (``+Lagrangian'')
reduces this to about $1.7$\,ms.

\Cref{tab:eval-planner-overhead} extends this across $nprobe$ and $K_{\max}$.
At common settings with $nprobe\le 256$, the planner runs in
$0.5\text{--}2.2$\,ms with all optimizations enabled, compared with
end-to-end query latencies of hundreds of milliseconds.
For settings where the unoptimized planner finishes within the
120-second limit, we
use its exact $T_P$ as ground truth. No optimized plan is more than
$3.5\%$ from that optimum, and the average error stays below $1.2\%$ at
every setting.

\begin{table}[t]
\setlength{\abovecaptionskip}{2pt}
\setlength{\belowcaptionskip}{0pt}
\caption{\textbf{Planner Overhead (SIFT1M).} Per-query p95 planner time
on real probe lists. At $nprobe{=}512$, Plain DP exceeds the 120\,s
timeout, so speedups are lower bounds.}
\label{tab:eval-planner-overhead}
\centering
\small
\renewcommand{\arraystretch}{0.92}
\begin{tabular*}{\linewidth}{@{\extracolsep{\fill}}lrcccc@{}}
\toprule
$nprobe$ & $K_{\max}$ & \shortstack{Plain DP\\p95 ms} & \shortstack{Full Opt.\\p95 ms} & \shortstack{Speedup\\vs. Plain} & \shortstack{Error (\%)\\avg./max} \\
\midrule
  \multirow{3}{*}{128} & 32 & 24.61 & 0.87 & 28.2$\times$ & 0.77 / 3.22 \\
   & 64 & 10.02 & 0.54 & 18.5$\times$ & 0.55 / 2.86 \\
   & 128 & 32.30 & 0.50 & 64.4$\times$ & 0.59 / 1.75 \\
\addlinespace[0.6pt]
  \multirow{3}{*}{256} & 64 & 441.6 & 1.90 & 232$\times$ & 1.09 / 2.79 \\
   & 128 & 837.9 & 1.72 & 486$\times$ & 0.48 / 1.75 \\
   & 256 & 1368 & 2.17 & 630$\times$ & 0.38 / 1.22 \\
\addlinespace[0.6pt]
  \multirow{3}{*}{512} & 128 & $>$120s & 4.72 & >25450$\times$ & n/a \\
   & 256 & $>$120s & 7.17 & >16733$\times$ & n/a \\
   & 512 & $>$120s & 9.90 & >12123$\times$ & n/a \\
\bottomrule
\end{tabular*}
\end{table}

\begin{figure*}[!t]
  \centering
  \begin{minipage}[t]{0.485\textwidth}
    \centering
    \includegraphics[width=\linewidth]{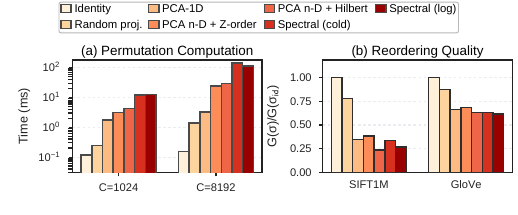}
    \captionof{figure}{\textbf{Reordering Method Comparison.} Left:
    single-threaded time to compute the cluster permutation. Right: normalized
    co-access gap on SIFT1M and GloVe. Both PCA $n$-D variants use $n{=}2$.}
    \label{fig:reorder-panels}
  \end{minipage}\hfill
  \begin{minipage}[t]{0.485\textwidth}
    \centering
    \includegraphics[width=\linewidth]{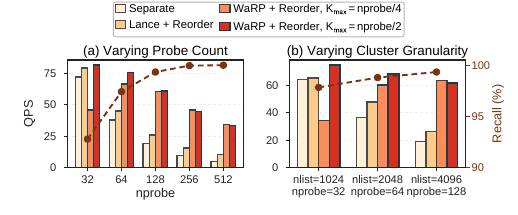}
    \captionof{figure}{\textbf{SIFT1M Probe Count and Cluster Granularity.}
    Left: $nlist{=}4096$, $nprobe \in \{32,64,128,256,512\}$. Right:
    $(nlist,nprobe)\in\{(1024,32),(2048,64),(4096,128)\}$ at fixed probe
    fraction.}
    \label{fig:eval-sensitivity-sift1m}
  \end{minipage}
  \Description{Two aligned two-panel figures. Figure 13 compares the
  computation time and normalized weighted co-access gap of seven reordering
  methods. Figure 14 varies probe count and cluster granularity on SIFT1M and
  compares Separate, Lance plus Reorder, and two WaRP plus Reorder request
  budgets, together with recall.}
\end{figure*}

\subsection{Reordering Method Comparison}
\label{sec:eval-reorder}

We compare the reordering methods introduced in
\Cref{sec:reorder-methods} based on their offline computation time and
the co-access locality they produce. \Cref{fig:reorder-panels} shows
the results. We measure computation time on cluster sets of size
$C \in \{1024,8192\}$, using random SIFT1M base vectors as simulated
centroids. We measure
locality on SIFT1M and GloVe using the
normalized weighted co-access gap
$G(\sigma)/G(\sigma_{\mathrm{id}})$ defined in
\Cref{sec:reorder-objective}. Here, $\sigma$ is the permutation produced
by a method and $\sigma_{\mathrm{id}}$ is the original cluster order, so
lower values indicate better co-access locality. We use $nlist{=}4096$
and $nprobe{=}128$, and fix both PCA $n$-D variants at $n{=}2$.

At the smaller $C{=}1024$ setting, every method finishes quickly. At
$C{=}8192$, closer to a production configuration, the spread widens.
The projection methods take
$1.4$--$30$\,ms while the spectral methods take $112$--$143$\,ms.
This extra spectral cost does not give consistently better locality.
PCA $n$-D + Hilbert performs best on SIFT1M, while Spectral (log)
performs best on GloVe. We deploy PCA $n$-D + \mbox{Z-order} because it needs
no query log, avoids the cubic cost of the spectral methods, and is easy
to extend to higher $n$.

\subsection{Probe Count and Cluster Granularity}
\label{sec:eval-scaling}

\system{} relies on merging adjacent reads, so its gain depends on
two properties of the IVF configuration, namely the size of each
cluster and the number of clusters probed per query. Both shift
with deployment choices in practice. Vector quantization at
different compression levels changes per-cluster payload, and
different recall targets push $nprobe$ up or down. We sweep these
two axes on SIFT1M, varying $nprobe$ at fixed $nlist$ for the
first, and raising $nlist$ and $nprobe$ together at fixed probe
fraction for the second. All runs use the reordered layout, query concurrency
$Q{=}4$, and an I/O concurrency limit of 64. \system{} + Reorder uses
request budgets $K_{\max}{=}nprobe/4$ and $K_{\max}{=}nprobe/2$.

In \Cref{fig:eval-sensitivity-sift1m}a, we sweep $nprobe$
from 32 to 512.
Separate issues one request per cluster, so its QPS declines sharply as
$nprobe$ increases. With $K_{\max}{=}nprobe/4$, \system{} + Reorder merges
the selected intervals, and its QPS declines more gradually. However, the
tight request budget requires more aggressive merging and leaves it behind
Separate at small $nprobe$. With $K_{\max}{=}nprobe/2$, \system{} + Reorder
stays above Separate throughout the sweep, and its advantage grows with
$nprobe$. In \Cref{fig:eval-sensitivity-sift1m}b, each query probes the
same fraction of the index as clusters become smaller. Separate declines by
$3.3\times$ as per-request
overhead dominates the smaller payloads, whereas \system{} + Reorder declines
by only $1.2\times$ because it merges the resulting small reads. Thus, its
advantage over Separate grows as cluster size decreases.

\section{Related Work}
\label{sec:related}

\textit{ANN Indexes.}
ANN indexes include tree-based
methods (KD-trees~\cite{KDTree}, RP-trees~\cite{RPTree}), hashing
(LSH~\cite{LSH1,LSH2}), partition methods (IVF/IVFPQ~\cite{ivfPQsift,faissivf},
IMI~\cite{IMI_babenko2014inverted}), and graph methods
(HNSW~\cite{HNSW}, NSG~\cite{NSG}). These indexes are often paired
with compression schemes such as PQ~\cite{ivfPQsift},
OPQ~\cite{OPQ}, and RabitQ~\cite{RabitQ}. Disk-aware variants
(DiskANN~\cite{diskann}, SPANN~\cite{chen_2021_sptag},
Starling~\cite{wang2024starling}, PipeANN~\cite{pipeannguo2025achieving},
PageANN~\cite{pageannkang2025scalable}) reduce random I/O on local
SSDs through graph layout and pipelining. Dynamic and filtered variants
have also been studied~\cite{freshdiskann,xu2023spfresh,
zhang2025cleann,ipdiskann,jiang2025digra}. \system{} instead operates
after IVF probe selection and optimizes retrieval of the selected clusters
from object storage.

\textit{Object Storage, Remote I/O, and Vector Retrieval.}
Within the same region, object-store reads incur per-request charges and
substantial fixed latency, making request granularity and concurrency central
to throughput. AnyBlob~\cite{Anyblob} empirically identifies a request-size
sweet spot for analytics, while Delta Lake~\cite{armbrust2020delta} highlights
broader object-store limitations, including costly metadata operations and
limited storage semantics.
\layoutmargin{27.72894pt}{5.98949pt}{2.64598pt}
\papertext{Vector systems use object storage at different points in the serving
path. Weaviate~\cite{weaviateIndexing2026} and
Qdrant~\cite{qdrantStorage2026} serve active indexes from memory or local
storage, while Milvus full-load mode~\cite{milvusArchitecture2026} caches
object-store-resident segments before search. Other designs access object
storage during query processing. Milvus Tiered
Storage~\cite{milvusTieredStorage2026} fetches data chunks or whole per-segment
index files on demand, while turbopuffer~\cite{turbopuffer2026architecture}
reads uncached data directly from object storage and caches it locally.
Such on-demand object-store access is useful when the full working set cannot
remain local or when loading an entire index before serving queries is
undesirable. For IVF indexes in this setting, \system{} reduces the number of
object-store requests by coalescing the byte intervals of cache-missing cluster
payloads into fewer Range-GETs.}
\layoutmargin{27.72894pt}{5.98949pt}{2.64598pt}
\papertext{On this query path, Lance~\cite{lance}
and the published Databricks design~\cite{databricks2025decoupled} expose
fixed gap or span thresholds for merging adjacent ranges. \system{} instead
formulates request merging as a global optimization problem and provides an
efficient planner. Our microbenchmarks in \Cref{fig:eval-qps-cost} compare
\system{} with this fixed-threshold family and show that \system{} + Reorder
achieves the highest peak QPS across all six datasets.}

\textit{Data Reordering and Locality Optimization.}
Space-filling curves such as Hilbert~\cite{kamel1994hilbert} and
Z-order curves~\cite{orenstein1986spatial} have long been used in databases to
linearize multidimensional data while preserving spatial
locality~\cite{hilbert1}. Modern data lake systems such as Apache
Hudi~\cite{hudi_zorder2021} apply Z-order and Hilbert ordering to improve
data skipping and reduce I/O for analytical queries. In the disk-based ANN
setting, DiskANN++~\cite{ni2023diskann++} and Starling~\cite{wang2024starling}
reorder graph vertices within fixed-size storage pages so that each page read
contains more useful neighbors. \layoutmargin{20.89795pt}{5.98949pt}{2.64598pt}
\papertext{\system{} instead reorders variable-size IVF cluster payloads in a
flat object-store layout to reduce the gaps among payloads likely to be probed
together. The online planner can then choose variable-length Range-GETs for
each query. \system{} therefore optimizes the layout for object-store request
planning rather than for a fixed-size read unit.}\looseness=-1

\section{Conclusion}
\label{sec:conclusion}
We present \system{}, a planner that optimizes how a vector index
issues requests to object storage. It uses a request-time surrogate
to estimate each request's completion time and an optimized planning
algorithm that is fast enough to run on the query path.
\layoutmargin{25.52394pt}{15.10497pt}{2.64598pt}
\papertext{We also introduce offline payload reordering to improve
co-access locality and create more merge opportunities.}
\papertext{Across seven datasets, \system{} + Reorder delivers
$1.09\text{--}3.48\times$ Lance's QPS and a
$2.12\text{--}5.19\times$ GET reduction.}
\papertext{Together, these results establish request planning as a
practical optimization layer between ANN probe selection and
object-store reads.}

\begin{acks}
This work was supported by the National Natural Science Foundation of China (Grant No. 62532001), Shanghai Qizhi Institute, and Tsinghua University - Keystone Electrical(Zhejiang) Co., Ltd Joint Research Center for Embodied Multimodal Artificial Intelligence (JCEMAI).
\end{acks}

\bibliographystyle{ACM-Reference-Format}
\bibliography{bib/references}

\end{document}